\documentclass[usenatbib,iop,numberedappendix]{aeb_emulateapj_2010}
\usepackage{amsmath}
\usepackage{amssymb}
\usepackage{xspace}
\usepackage[normalem]{ulem}
\usepackage{mathrsfs}

\usepackage{natbib}

\def\del#1{{}}

\def\s{{\rm s}} 

\def\m{{\rm m}} 
\def\cm{{\rm c}\m} 

\def\eV{{\rm eV}} 
\def\GeV{{\rm G}\eV} 

\newcommand{\rmn}{\mathrm}

\newcommand\bmath[1] {\mbox{\boldmath$\rm #1$}}

\def\N{\mathcal{N}}
\def\F{\mathcal{F}}
\def\Fermi{{\em Fermi}\xspace}

\begin{document}

\title{
Lower Limits on the Anisotropy of the Extragalactic Gamma-Ray
Background\\
implied by the 2FGL and 1FHL Catalogs
}

\author{
Avery E.~Broderick\altaffilmark{1,2},
Christoph Pfrommer\altaffilmark{3},
Ewald Puchwein\altaffilmark{3},
Philip Chang\altaffilmark{4}, and
Kendrick M.~Smith\altaffilmark{1}
}
\altaffiltext{1}{Perimeter Institute for Theoretical Physics, 31 Caroline Street North, Waterloo, ON, N2L 2Y5, Canada}
\altaffiltext{2}{Department of Physics and Astronomy, University of Waterloo, 200 University Avenue West, Waterloo, ON, N2L 3G1, Canada}
\altaffiltext{3}{Heidelberg Institute for Theoretical Studies, Schloss-Wolfsbrunnenweg 35, D-69118 Heidelberg, Germany}
\altaffiltext{4}{Department of Physics, University of Wisconsin-Milwaukee, 1900 E. Kenwood Boulevard, Milwaukee, WI 53211, USA}

\shorttitle{Limits on the EGRB Anisotropy Spectrum }
\shortauthors{Broderick et al.}

\begin{abstract}
  \noindent
  In principle, the angular anisotropy in the extragalactic gamma-ray
  background (EGRB) places severe constraints upon putative
  populations of unresolved gamma-ray point sources.  Existing
  estimates of the EGRB anisotropy have been constructed by excising
  known point sources, e.g., taken from the First or 2 Year \Fermi-LAT
  Source Catalog (1FGL or 2FGL, respectively) and statistically
  analyzing the residual gamma-ray sky maps.  We perform an
  independent check of the EGRB anisotropy limits by comparing the
  values obtained from the 1FGL-masked sky maps to the signal implied
  by sources that lie below the 1FGL detection threshold in the
  more sensitive 2FGL and 1FHL (First \Fermi-LAT catalog of $>10$~GeV
  sources). As such, our analysis provides an internal consistency
  check of implications for source counts and spectral index
  distributions of gamma-ray bright active galactic nuclei obtained
  from \Fermi-LAT data.  Based on this, we find evidence for
  substantially larger anisotropies than those previously reported at
  energies above 5~GeV, where BL Lac objects are likely to provide
  the bulk of their contribution to the EGRB.  This uncertainty in
  the EGRB anisotropy cautions against using it as an independent
  constraint for the high-redshift gamma-ray universe.  Moreover, this
  would suggest that contrary to previous claims, smooth extensions of
  the resolved point-source population may be able to simultaneously
  explain both the isotropic and anisotropic components of the EGRB.
\end{abstract}


\maketitle

\section{Introduction} \label{sec:I}
The Large Area Telescope (LAT) on board the \Fermi space telescope has
proven to be a powerful tool for studying the extragalactic component
of the gamma-ray sky.  It has already detected more than 1800 discrete
point sources, the overwhelming majority of which is associated with
active galactic nuclei (AGNs; specifically blazars), with small populations
of dimmer, though potentially much more numerous, sources
\citep[e.g., starburst and radio galaxies, see Table 6 of][]{2FGL}. 
These observations have produced the strongest constraints to date
upon the gamma-ray bright blazar population and its evolution.

Likewise, the unresolved component of the extragalactic gamma-ray sky
constrains the nature and number
of point sources that lie below the current detection threshold.  At
present this arises in two contexts: (1) the isotropic extragalactic
gamma-ray background (EGRB) spectrum, which sets the total flux that
must be accounted for, and (2) the angular structure in the EGRB,
characterized by an anisotropy coefficient (see below), that limits
the fraction of the EGRB that can be produced by bright point sources.
Of these, the second has the most severely constrained models that produce
the EGRB from extensions of the gamma-ray point-source populations to
low fluxes and high redshifts.

Attempts to model the EGRB by extending existing point-source
populations receive strong theoretical support from our current
understanding of the history of baryon accretion into dark matter
halos and its associated observational tracers.  Such models have
been generally successful in reproducing the observed EGRB
brightness and energy spectrum, both via phenomenological and
physically motivated extensions \citep[see,
e.g.,][]{SPA12,PaperI,PaperV,Cavadini+2011}.  They also receive
qualitative support from the structure of the measured EGRB angular
power spectrum, $C_\ell$, within a variety of energy bands. Between
$\ell=150$ and 500, the $C_\ell$ are constant, consistent with the EGRB
being produced by a population of discrete sources with angular scales
smaller than the width of the \Fermi point-spread function
\citep[PSF;][]{Fermi_aniso}.

Quantitatively, however, the EGRB power spectrum is {\em not}
consistent with point-source populations that are extensions of the
known gamma-ray bright source populations.  Studies that have
purported to show such agreement have done so accounting for only a
subset of the known point-source population \citep[see,
e.g.,][]{Cuoco,Harding:2012}.  The origin of the limit is easy to
understand: if the EGRB is produced by a few sources just below the
current detection thresholds, it should exhibit a correspondingly
large degree of angular variability.  From this, it has been argued
that no more than 20\% of the isotropic EGRB can be due to gamma-ray
blazars without violating the EGRB anisotropy constraints, and thus
motivated renewed interest in alternative gamma-ray sources
\citep{Cuoco}.  However, this conclusion is predicated upon the proper
normalization of the angular power.

This normalization remains uncertain for at least two reasons.  First,
it is necessary to deconvolve the \Fermi-LAT PSF from the observed
EGRB power spectrum to obtain the desired intrinsic spectrum, from
which the EGRB anisotropy is determined.  As a result, any errors in
the assumed PSF will inevitably generate corresponding errors in the
anisotropy normalization.  Second, the uncertainty estimates
were based upon expressions that assumed the background is well
approximated by a Gaussian random field.  While these are appropriate
for large populations of intrinsically weak sources, they are
misleading when the EGRB anisotropy is dominated by a handful of point
sources near the detection threshold.  In the latter case, the observed
gamma rays are strongly correlated via the nature of their origin,
resulting in a substantially larger EGRB anisotropy uncertainty.

\begin{deluxetable*}{lccccc}[th!]
\tablecaption{Estimated Lower Limits upon the EGRB Anisotropy\tablenotemark{a} \label{tab:CPs}}
\tablehead{
& & \multicolumn{4}{c}{$\Delta C_P$ within various energy bands ($10^{-19}~{\rm ph^2 cm^{-4} s^{-2} sr^{-1}}$)}
\\
&&&&&\\
Method\tablenotemark{b}
&
Section
&
1.04--1.99~GeV
&
1.99--5.00~GeV
&
5.00--10.4~GeV
&
10.4--50~GeV
}
\startdata
\citet{Fermi_aniso}, {\sc data:cleaned} & --- & $46.2\pm11.1$ & $13.0\pm2.2$ & $0.85\pm0.25$ & $0.211\pm0.086$\\
\hline\\
Unmasked 2FGL detected in each energy band & 3.1 & $27.5\pm~~5.0$ & $~~7.0\pm1.3$ & $0.63\pm0.13$ & $0.168\pm0.063$\\
Unmasked 1FHL detected above 10~GeV & 3.1 & --- & --- & --- & $0.238\pm0.092$\\
Unmasked 2FGL band-corrected $\F_{35}$  & 3.2 & $26.9\pm~~3.7$ & $~~8.6\pm1.1$ & $0.86\pm0.11$ & $0.54~~\pm0.09~~$\\
\hline\\
Statistical 2FGL, w/2FG & 4.3 & $38.5\pm~~2.8$ & $12.5\pm0.9$ & $1.35\pm0.11$ & $0.98~~\pm0.12~~$\\
Statistical 2FGL, w/2FGL-EB & 4.3 & $16.0\pm~~2.7$ & $~~5.2\pm0.9$ & $0.61\pm0.11$ & $0.52~~\pm0.10~~$\\
Statistical 2FGL, w/PL & 4.3 & $49.8\pm~~3.4$ & $16.1\pm1.0$ & $1.71\pm0.13$ & $1.20~~\pm0.13~~$\\
Statistical 2FGL, w/PL-EB & 4.3 & $36.7\pm~~2.0$ & $11.9\pm0.6$ & $1.29\pm0.08$ & $0.94~~\pm0.10~~$\\
Statistical 2FGL, hard sources w/2FGL & 4.3 & $~~6.9\pm~~0.9$ & $~~4.1\pm0.5$ & $0.73\pm0.09$ & $0.76~~\pm0.11~~$\\
Statistical 2FGL, hard sources w/2FGL-EB & 4.3 & $~~3.5\pm~~0.7$ & $~~2.1\pm0.5$ & $0.39\pm0.09$ & $0.45~~\pm0.10~~$\\
Statistical 2FGL, hard sources w/PL & 4.3 & $~~8.9\pm~~1.1$ & $~~5.1\pm0.6$ & $0.90\pm0.10$ & $0.91~~\pm0.13~~$\\
Statistical 2FGL, hard sources w/PL-EB & 4.3 & $~~7.0\pm~~0.8$ & $~~4.0\pm0.4$ & $0.71\pm0.07$ & $0.73~~\pm0.10~~$\\
\hline\\
Statistical power law extension w/$\F_{35}>3\times10^{-10}~{\rm ph~cm^{-2} s^{-1}}$ and PL & 4.4 & $84.9\pm~~9.2$ & $27.2\pm2.8$ & $2.80\pm0.34$ & $1.73~~\pm0.30~~$\\
Statistical power law extension w/$\F_{35}>3\times10^{-10}~{\rm ph~cm^{-2} s^{-1}}$ and PL-EB & 4.4 & $68.5\pm~~5.9$ & $21.9\pm1.8$ & $2.26\pm0.22$ & $1.40~~\pm0.19~~$\\
Statistical power law extension w/$\F_{35}>10^{-10}~{\rm ph~cm^{-2} s^{-1}}$ and PL & 4.4 & $145.0\pm~~9.6~~$ & $46.4\pm3.0$ & $4.78\pm0.35$ & $2.96~~\pm0.31~~$\\
Statistical power law extension w/ $\F_{35}>10^{-10}~{\rm ph~cm^{-2} s^{-1}}$ and PL-EB & 4.4 & $128.6\pm~~6.6~~$ & $41.2\pm2.0$ & $4.24\pm0.24$ & $2.63~~\pm0.21~~$\\
Statistical power law extension w/$\F_{35}>10^{-12}~{\rm ph~cm^{-2} s^{-1}}$ and PL & 4.4 & $187.4\pm~~9.6~~$ & $60.0\pm3.0$ & $6.18\pm0.35$ & $3.83~~\pm0.31~~$\\
Statistical power law extension w/$\F_{35}>10^{-12}~{\rm ph~cm^{-2} s^{-1}}$ and PL-EB & 4.4 & $171.1\pm~~6.6~~$ & $54.8\pm2.0$ & $5.64\pm0.24$ & $3.50~~\pm0.22~~$
\enddata
\tablenotetext{a}{The $1\sigma$ uncertainty intervals indicate the cosmic variance and include the propagation of flux uncertainties.}
\tablenotetext{b}{See Table \ref{tab:wfits} for the definitions of the threshold functions 2FGL, 2FGL-EB, PL, and PL-EB.}
\end{deluxetable*}

In addition, the general success of extensions of the known blazar
population to low fluxes provides a natural reason to revisit the EGRB
anisotropy normalization.  However, more disturbing is the precipitous
decline in blazar numbers below the \Fermi detection threshold
required for consistency.  This is difficult to envision, even in
principle, requiring a pathological redshift evolution and luminosity
function for which there is currently no other evidence 
\citep[][see also Section~\ref{sec:threshold} of the present work]{Harding:2012}. 
Most disconcerting, this would represent an obstacle to unifying the
gamma-ray blazar population with that of other AGNs, would be at odds
with the underlying physical picture of accreting black hole systems,
and would suggest a conspiracy between accretion physics and the
formation of structure. By analogy with Galactic X-ray binaries, in
the unified picture of AGNs, these systems are thought to switch
between their accretion states, cycling between the high/soft state
(i.e., quasars) and the low/hard state (i.e., radio-loud AGNs) on the
dynamical timescale of the inner accretion flow, instead of those
associated with the supply of available gas
\citep{Macc-Gall-Fend:03,McHardy:2006}. This picture would predict
that AGN types associated with the various accretion states to have
contemporaneous populations. This receives indirect support from the
apparent rapid redshift evolution of radio-loud AGNs, the presumed
parent population of blazars, which peak near $z\sim1.85$
\citep{Will_etal:01,Wall:2005}.

Motivated by this, here we present an external, empirical assessment
of the normalization of the reported EGRB anisotropy.  We do this by
exploiting the sensitivity difference between the First and 2 Year
\Fermi-LAT Source Catalogs \citep[1FGL and 2FGL,
respectively,][]{1FGL,2FGL} and the more recently published First
\Fermi-LAT catalog of $>10$~GeV sources \citep[1FHL,][]{1FHL}.  The
EGRB anisotropy measurements reported by \citet{Fermi_aniso} employed
the 1FGL for point-source identification and removal.  Despite this,
the \Fermi data used for the anisotropy measurement were collected over
a 22 month period, coincident with the 24 month period used to
construct the more sensitive 2FGL and 36 month period used to
construct the 1FHL.  Thus, subsets of the 2FGL and 1FHL
will have contributed to the observed EGRB anisotropy calculated
  from sky maps where 1FGL point sources have been masked. By
comparing their contributions to the observed EGRB anisotropy with the
reported values, we obtain independent limits on the EGRB anisotropy
magnitude.

Note that this is quite different than the discussion of the
dependence upon the catalog (1FGL or 2FGL) used to construct the
point-source mask in \citet{Fermi_aniso}. There the fraction of the
EGRB anisotropy associated with the 2FGL was assessed by comparing
the anisotropy signal in residual sky maps obtained by masking on
the 2FGL sources instead of the 1FGL (which comprised their main
analysis). Here, we are concerned with verifying the absolute
normalization of the anisotropy signal using an {\em independent}
and potentially more robust method that is based directly on the
properties of the unmasked point sources listed in the 2FGL and 1FHL
rather than on the angular power spectra of the residual sky maps.

In the interests of completeness, we present a variety of estimates of
the expected anisotropy.  These may be broadly placed into two
categories: estimates that explicitly use the sources in the 2FGL and
1FHL that are unmasked by the 1FGL (which we call ``direct
estimates'') and estimates that use the entire 2FGL and 1FHL as
statistical measures of the unmasked source population (``statistical
estimates'').  The former set is much more susceptible to
cosmic variance, though places hard lower limits on the measured
EGRB anisotropy spectrum.  In practice, these are almost certainly
substantial underestimates, as they are predicated upon the physically
unreasonable assumption that sources below the band-specific detection
threshold contribute nothing.  The statistical estimates are likely
more robust as a result of the considerably larger statistical sample
from which they are drawn.  Perhaps more importantly, they are more
directly relevant for comparison to theoretically motivated source
populations, which typically assume isotropy, and are thus insensitive
to the potentially large cosmic variance associated with the
considerably smaller masked population.  The assumption made is that
an accurate statistical description of the sources that remain after
masking on the 1FGL can be obtained from the 2FGL source population
(which we will justify empirically).  In all cases, we find
qualitatively similar results: in the absence of a pathological
decline in high-energy gamma-ray flux in sources just below the
\Fermi~detection threshold at high energies, the reported EGRB
anisotropy calculated from masking the 1FGL point sources is
significantly below what may be already accounted for from the
observed point-source populations in the 2FGL and 1FHL.  A summary
of the lower limits upon the anisotropy within the energy bands reported in
\citet{Fermi_aniso} is collected in Table \ref{tab:CPs}, together with
a brief description of the method employed and the section where it is
described.

We begin in Section \ref{sec:CP} with a short derivation of the
contribution to the EGRB anisotropy spectrum from a population of
point sources and a description of how we construct the relevant
fluxes for each band.  We present direct estimates for various 
subpopulations of the 2FGL and 1FHL in Section \ref{sec:direct}.
Statistical estimates are collected in Section \ref{sec:statistical}.
In Section \ref{sec:discussion}, we discuss our results in the context
of the measured EGRB and possible reasons for differences.  Finally,
conclusions are collected in Section \ref{sec:conclusions}.

\section{Constructing Anisotropy Power Spectra}\label{sec:CP}

\subsection{Definition of $C_P$}
Above $\ell\simeq150$, the EGRB angular power
spectrum is well characterized by a constant, i.e., $C_\ell = C_P$ \citep{Fermi_aniso}.
In the flat-sky limit,\footnote{
Expanding point-source sky maps obtained in Section \ref{sec:direct}
into spherical harmonics, we measure the angular power spectrum and
verify that it becomes flat for multipoles $\ell>15$ and 
consistent with a Poisson power spectrum.  This provides an explicit
validation of the flat-sky limit at multipoles of interest.} this is
related to the gamma-ray flux map via 
\begin{equation}
C_\ell
=
\int d^2\theta e^{i\bmath{\ell}\cdot\bmath{\theta}}
\int \frac{d^2\theta'}{4\pi} \F(\bmath{\theta}'+\bmath{\theta}) \F(\bmath{\theta}')\,,
\end{equation}
where for a given ensemble of unresolved point sources with
fluxes $\{\F_j\}$ and positions $\bmath{\theta}_j$ the flux per unit
solid angle is 
\begin{equation}
\F(\bmath{\theta}) = \sum_j \F_j \delta^2(\bmath{\theta}-\bmath{\theta}_j)\,.
\end{equation}
Inserting this into the above equation and averaging over ensembles (including
source sky positions) yields
\begin{equation}
\label{eq:Cell}
\begin{aligned}
\left< C_\ell \right>
&=
\left<
\int d^2\theta e^{i\bmath{\ell}\cdot\bmath{\theta}}
\int \frac{d^2\theta'}{4\pi} \sum_{j,k}\F_j\F_k
\delta^2(\bmath{\theta}'+\bmath{\theta}-\bmath{\theta}_j)
\delta^2(\bmath{\theta}'-\bmath{\theta}_k)
\right>\\
&=
\left<
\sum_{j,k}
\frac{\F_j \F_k}{4\pi}
e^{i\bmath{\ell}\cdot\left(\bmath{\theta}_j-\bmath{\theta}_k\right)}
\right>\\
&=
\left< \sum_j \frac{\F_j^2}{4\pi} \right>\,,
\end{aligned}
\end{equation}
where we identify the final expression with $C_P$. Note that
larger individual fluxes or an enhanced number of sources increases
this dimensional measure of the anisotropy coefficient $C_P$.

\subsection{Relationship to Source Distributions}
If the flux distribution of {\em unresolved} sources is a function of
$\F$ only, this reduces to the standard expression
$C_P = \int d\F\,\F^2\,d\N/d\F$, where $d\N/d\F$ is the number of
sources {\em unresolved} in the 1FGL per steradian per unit flux.
Generally, this can be quite complicated, with the result that
\begin{equation}
C_P
=
\int d^n\!p\,\F^2\frac{d\N}{d^n\!p}\,,
\label{eq:CPdef}
\end{equation}
where $\bmath{p}$ are some set of $n$ parameters needed to describe
the flux distribution (e.g., flux and spectral index, or in our case
here, the 1--100~GeV band flux $\F_{35}$ and spectral index).
Further progress requires an explicit estimate for $d\N/d^n\!p$.

Here, we will use the 2FGL and 1FHL to provide an approximation of the
source population that is (statistically) not detected in the
1FGL. We do this primarily via distributions of the form
\begin{equation}
\frac{d\N}{d\F d\Gamma} \simeq \sum_j \frac{w_j}{\Omega_{\rm sky}}
\delta(\F-\F_j) \delta(\Gamma-\Gamma_j)\,,\label{eq:dNdFdGamma}
\end{equation}
where $\Omega_{\rm sky}$ is the fraction of the sky being included,
$F_j$ are the band-specific fluxes, and the $w_j$ are source-dependent
weights describing the likelihood that a given source was not
observed in the 1FGL\footnote{Modulo the weights, $w_j$, this is simply the
  approximation of the probability distribution often used for
  bootstrap methods.}.
Effectively, in Equation (\ref{eq:dNdFdGamma}), we are constructing
the source distribution directly from the 2FGL, as opposed to
utilizing fits to previously selected forms of $d\N/d\F d\Gamma$, as
has been the case elsewhere \citep[see, e.g.,][]{Cuoco}.
These empirical estimates of the source distribution correspond to
conservative estimates of the lower limit on the EGRB anisotropy from
unresolved blazars, with objects unresolved in the 2FGL further
enhancing the anisotropy.
For comparison, we also consider a simple
extrapolation of the 2FGL $d\N/d\F d\Gamma$ to assess the
fraction of the contribution to the EGRB anisotropy arising from
sources in the 2FGL alone.

Inserting Equation (\ref{eq:dNdFdGamma}) into the estimate for $C_P$,
Equation (\ref{eq:CPdef}), gives our conservative estimate of the 2FGL
contribution to the measured EGRB anisotropy:
\begin{equation}
C_P = \frac{1}{\Omega_{\rm sky}} \sum_j \F_j^2~w_j\,.
\label{eq:CP}
\end{equation}
This procedure is then repeated for each energy band reported in
\citet{Fermi_aniso}, yielding a construction of the $C_P$ spectrum.

\subsection{Cosmic Variance Error Estimate}
While a number of potential sources of uncertainty are present in
principle, after the detection efficiency, chief among them is cosmic
variance.  We define this to be the statistical uncertainty associated with
the choice of a particular realization of point sources from the
given flux distribution, which remains even after the individual
source parameters are fully characterized.  Via a procedure similar
to that employed above, we estimate this to be
\begin{equation}
\label{eq:sigma_Cp}
\sigma_{C_P}
=
C_P
\frac{
\sqrt{\sum_j \F_j^4 w_j}}{\sum_j \F_j^2 w_j}
\,.
\end{equation}
Typical values of $\sigma_{C_P}/C_P$ are 6\%--10\%.  By comparison,
the error induced by the intrinsic uncertainty in the band-corrected
flux, ignoring any uncertainty associated with the spectral shape
correction (see Appendix \ref{sec:K-shape-corr}), is estimated to be
roughly 2\%--7\% (see Appendix \ref{sec:error_prop}).  Where
relevant, both are included in the error estimates shown.  Note that
because the 2FGL contribution to the EGRB anisotropy is constructed
directly from the point sources, there is no photon noise term (beyond
that associated with the point-source flux uncertainties themselves).

To make further progress, we must describe how the
weights, $w_j$, are determined and how estimates for the band-specific
fluxes, $\F_j$, are obtained.
The remainder of this section describes how the band-specific
fluxes are obtained.  We defer a full discussion of the weights until
the presentation of the direct (Section (\ref{sec:direct}) and
statistical (Section \ref{sec:statistical}) approaches to the
construction of the source distribution.  

\subsection{Critical Importance of Spectral Information}
\begin{figure}
\begin{center}
\includegraphics[width=0.9\columnwidth]{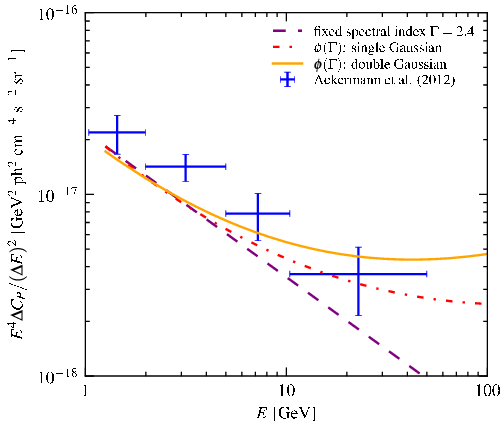}
\end{center}
\caption{Reported EGRB anisotropy spectrum 
  \citep[blue error bars, taken from][]{Fermi_aniso} in comparison to
  that implied by the flux distribution presented in \citet{Cuoco} for
  various choices of the photon spectral index distribution.  Shown in
  black is the model presented in \citet{Cuoco}, in which the photon
  spectral index is fixed at the average value for blazars; in red,
  a single-Gaussian model based upon the total blazar population from
  \citet{Fermi_lNlS}; in purple, a double-Gaussian model based upon
  separate FSRQ and BL Lac populations.  As the hard population is
  modeled with increasing accuracy, the anisotropy at high energies
  rises substantially, inducing a typically concave shape.  Note that
  the \citet{Cuoco} flux distribution underestimates the total number
  of 2FGL sources near the flux threshold by roughly a factor of two.}
\label{fig:GaussianCPs}
\end{figure}
Suggested by the choice of parametrization in the previous section,
the distribution with respect to the spectral shape is critical to
accurately estimating $C_P$.  While this is briefly alluded to in
\citet{Cuoco}, here we emphasize this point.  That is, power-law
anisotropy spectra are a result of pathological photon spectral index
distributions and are generally {\em not} expected.  Commonly, the
paucity of hard sources combined with their increasing dominance of
the background at high energies results in a characteristic concave
anisotropy spectrum, with potentially large enhancements above
10~GeV.  While some spectral softening does occur above 30~GeV (see
Appendix \ref{sec:K-shape-corr}), potentially leading to a decrease in
$C_P$ at high energies, reproducing a power-law anisotropy spectrum
requires an unlikely conspiracy between the spectral index
distribution at low energies and the softening at high energies.

Figure \ref{fig:GaussianCPs} shows the anisotropy spectrum obtained
for the flux distribution employed in \citet{Cuoco} for various
spectral index distributions.\footnote{Note that the fitted 1LAC flux
  distribution \citet{Cuoco} employed is shown in comparison to the
  observed 2FGL flux distribution in Figure \ref{fig:cut}, and
  underpredicts the number of sources by roughly a factor of two near
  the detection threshold.  We consider it here for comparison only.}
Effectively, these modify the band correction that is used to relate
the anisotropy at different energy bands.  For a single power-law
spectrum, the photon flux between energies $E_m$ and $E_M$ is related to
the photon flux between $E_x$ and $E_y$ via a band correction, i.e.,
\begin{equation}
  \F_{mM,j} = K(\Gamma_j) \F_{xy,j}
  \quad\text{where}\quad
  K(\Gamma) \equiv
  \frac{E_m^{1-\Gamma}-E_M^{1-\Gamma}}{E_x^{1-\Gamma}-E_y^{1-\Gamma}}\,.
\label{eq:Kdef}
\end{equation}
In the limit that $E_M-E_m\ll E_m,\,E_M$, the band correction becomes
\begin{equation}
K(\Gamma,E) \equiv (1-\Gamma) \Delta E \frac{E^{-\Gamma}}{E_x^{1-\Gamma}-E_y^{1-\Gamma}}\,,
\end{equation}
in terms of which
\begin{equation}
C_P(E) = C_{P,xy} \int d\Gamma  K(\Gamma,E)^2 \phi(\Gamma)\,,
\label{eq:CPspec}
\end{equation}
where $\phi(\Gamma)$ is the normalized photon spectral index distribution.

In their anisotropy estimates, \citet{Cuoco} assumed a fixed value of
$\Gamma=2.40$, corresponding to $\phi(\Gamma)=\delta(\Gamma-2.40)$.
This is shown by the purple long-dashed line in Figure
\ref{fig:GaussianCPs} for the flux distribution employed there.
However, in practice, this almost certainly significantly
underestimates the anisotropy at high energies, which is dominated by
the subset of hard sources.  This is evident by the remaining lines,
differing only in their choice of the spectral index distribution.

In particular, in Figure \ref{fig:GaussianCPs}, we also show models
with $\phi(\Gamma)$ given by a single Gaussian and a double Gaussian
with equal weights, meant to reproduce the intrinsic spectral
distributions found for all blazars and for the FSRQs and BL Lac objects
separately by \citet{Fermi_lNlS}.\footnote{We use the 1LAC sample
properties here for consistency with \citet{Cuoco}.}  In the
single-Gaussian model the mean and standard deviation are 2.4 and
0.24, respectively, which is sufficient to substantially bend the anisotropy
spectrum upward at high energies.  This is even more explicit for the
double-Gaussian model, which models the softer FSRQs (mean 2.47 and
std.~dev.~0.19) and harder BL Lac objects (mean 2.20 and std.~dev.~0.22)
separately.  Since this increases the low-$\Gamma$ wing,
it further enhances the high-energy anisotropy.

While a seemingly minor distinction, it is clear from 
Equation (\ref{eq:CPspec}) that it is the mean $K^2$ that enters into
the definition of $C_P(E)$, not the mean $\Gamma$ or mean $K$.
Nevertheless, this has significant implications where the source
population is dominated by a small number of objects.
For these reasons, we take special care to ensure that the impact of
the spectral distribution is accounted for via variations upon
Equation (\ref{eq:dNdFdGamma}).  Because of this, in all of the
anisotropy spectra presented here, the associated concave behavior is
readily apparent as anticipated.

\subsection{Summary of Band Correction and the Impact of Spectral Softening}

\begin{figure}
\begin{center}
\includegraphics[width=0.8\columnwidth]{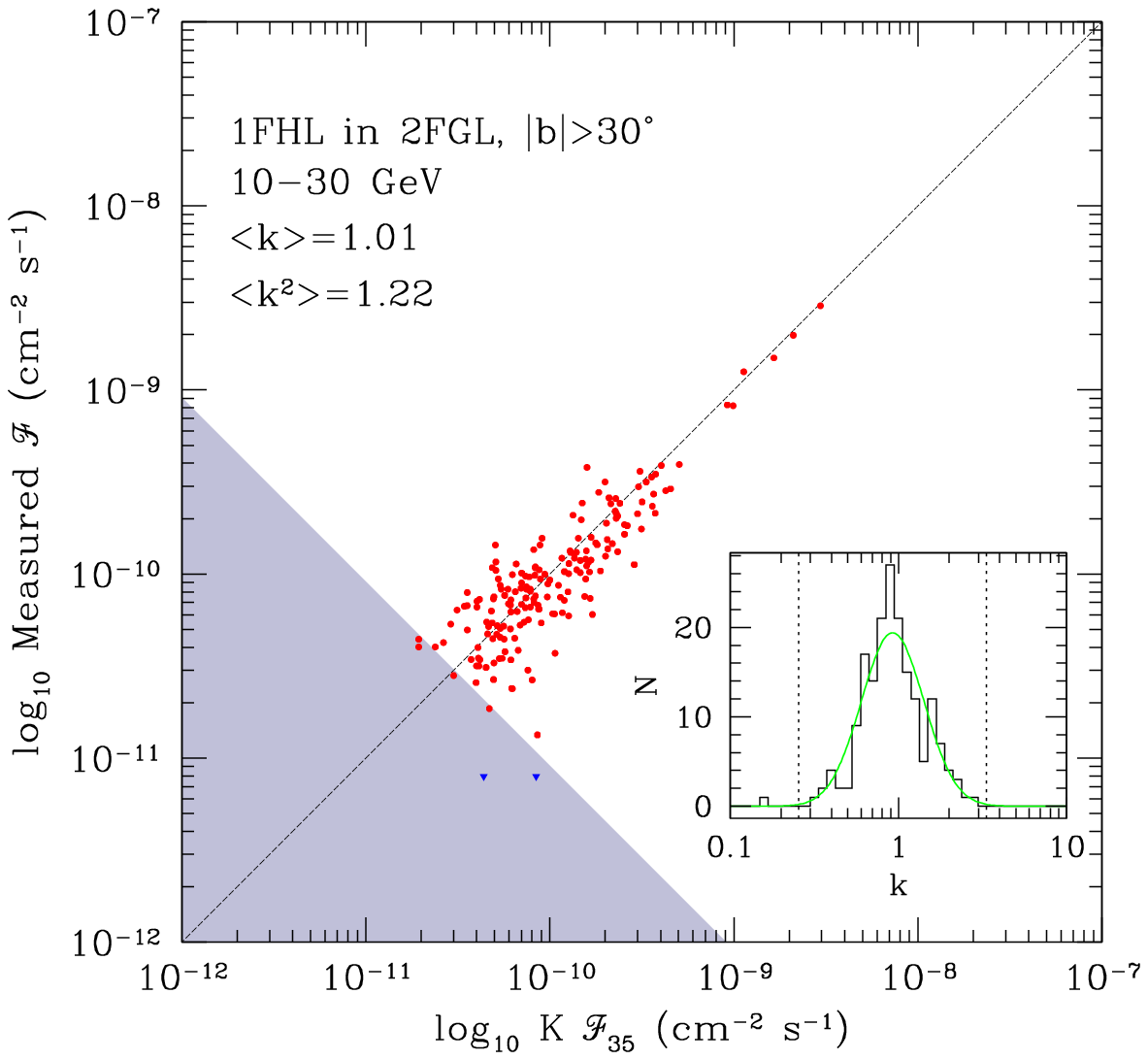}
\end{center}
\caption{Comparison of the 10~GeV--30~GeV flux reported in the 1FHL
  and obtained by a band correction from the 1~GeV--100~GeV 2FGL flux
  (assuming power-law spectra) for sources that appear in both
  catalogs.  Measured fluxes are shown by the red squares; upper
  limits are shown by the blue triangles. The detection threshold is
  denoted by the grayed region in the lower left.  For reference, the
  dotted line shows equality.  Spectral softening, which we ultimately
  account for via the spectral shape correction, $k_{mM}$, is
  responsible for the offset between the observed fluxes and those
  obtained by the band correction alone.  Shown in the inset is the
  distribution of the spectral shape correction.  See Appendix
  \ref{sec:K-shape-corr} for more detail regarding how this is derived
  and the average values constructed.  (Note that while all sources in
  the 1FHL are detected above 10~GeV, not all 1FHL sources are
  detected in the specific 10~GeV--30~GeV band shown here.)
}\label{fig:kcorr_1FHL}
\end{figure}

\begin{figure}
\begin{center}
\includegraphics[width=0.8\columnwidth]{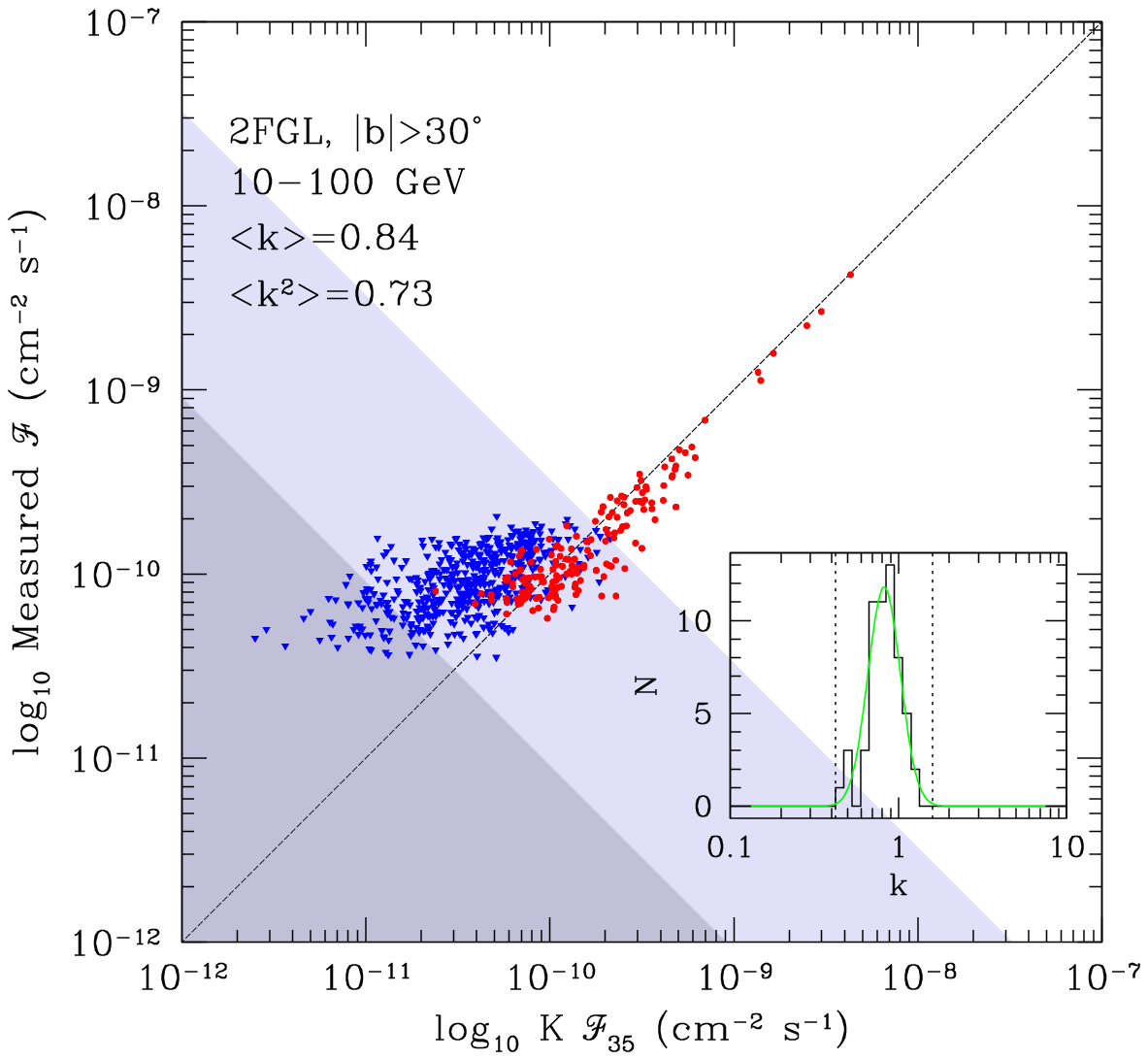}
\end{center}
\caption{Comparison of the 10~GeV--100~GeV flux reported in the 2FGL
  and obtained by a band corr1ection from the 1~GeV--100~GeV 2FGL flux
  (assuming power-law spectra).  Measured fluxes are shown by the red
  squares; upper limits are shown by the blue triangles.  The 2FGL and
  1FHL detection thresholds are denoted by the light and dark gray
  regions in the lower left, respectively.  For reference, the dotted
  line shows equality.  Spectral softening, which we ultimately
  account for via the spectral shape correction, $k_{mM}$, is
  responsible for the offset between the observed fluxes and those
  obtained by the band correction alone.  Shown in the inset is the
  distribution of the spectral shape correction.  See Appendix
  \ref{sec:K-shape-corr} for more detail regarding how this is derived
  and the average values are constructed.}\label{fig:kcorr_2FGL}
\end{figure}

Where sources are detected within the particular energy bands in which
the anisotropy has been reported, no band correction is needed.
However, it is typically the case that source fluxes either are reported
only in nearby energy ranges or are not detected at all at high
energies.  For these cases, some band correction is required.  We
describe this process is detail in Appendix \ref{sec:K-shape-corr},
and thus summarize the salient points here.

In the former case, when the source is detected at high energies in
adjacent energy bands, we simply apply the power-law band correction
in Equation (\ref{eq:Kdef}).  The error from this procedure is
expected to be small because of the small extrapolation involved.

Nevertheless, when the source is detected only at low energies, the
applicability of the power-law band correction is not immediately
obvious.  We address that here by comparing the power-law band-corrected 1~GeV--100~GeV fluxes to measured sub-band fluxes for
sources within the 2FGL and sources that appear in both the 2FGL and
1FHL (the latter case providing better energy resolution above
10~GeV).  Examples may be found in Figures \ref{fig:kcorr_1FHL} and
\ref{fig:kcorr_2FGL}, as well as in Appendix
\ref{sec:K-shape-corr}. Note that we apply a condition solely upon
the photon flux from 1 to 100~GeV ($\F_{35}$) since there is no
systematic correlation between $\F_{35}$ and the photon spectral
index, $\Gamma$ (see Section~\ref{sec:threshold} and
\citealt{Cuoco}).

\begin{deluxetable}{ccccc}
\tablecaption{Spectral Corrections to the band-correction for \Fermi
  Point Sources. \label{tab:kcorr}}
\tablehead{
$E_m$\tablenotemark{a}
&
$E_M$\tablenotemark{a}
&
$F_{\rm min}$\tablenotemark{b}
&
$\left<k\right>$
&
$\left<k^2\right>$
}
\startdata
0.3 & 1   & $2\times10^{-9}$ &    1.01 & 1.05 \\
1   & 3   & $5\times10^{-10}$ &   1.06 & 1.16 \\
3   & 10  & $1.5\times10^{-10}$ & 1.04 & 1.13 \\
10  & 30  & $3\times10^{-11}$ &   1.01 & 1.22 \\
30  & 100 & $2\times10^{-11}$ &   0.88 & 1.08 \\
100 & 500 & $2\times10^{-11}$ &   0.47 & 0.29 
\enddata
\tablenotetext{a}{In units of GeV}
\tablenotetext{b}{In units of ${\rm cm^{-2}s^{-1}}$}
\end{deluxetable}

Below 30~GeV, the power-law band-corrected fluxes provide an excellent
fit to the measured fluxes.  At higher energies, the relationship
remains linear, with a constant of proportionality that decreases with
the energy band, indicating the presence of spectral softening (which
becomes substantial above 100~GeV).  For this reason, we define a
spectral shape correction factor
\begin{equation}
k_{mM} \equiv \frac{\F_{mM}}{K \F_{35}}\,.
\end{equation}
Within each energy band, the $k_{mM}$ appear well characterized by a
log-normal distribution, with a width that is independent of source
flux and increases with gamma-ray energy.  We therefore assume that
the $k_{mM}$ are log-normal random variables, in terms of which the
reconstructed flux is $\F_{mM}=k_{mM} K\F_{35}$.  The mean $k_{mM}$,
$\left< k\right>$, are listed in Table \ref{tab:kcorr} and are
comparable to unity below 30~GeV, falling to $0.88$ for
30~GeV--100~GeV, and dropping rapidly thereafter, presumably due
to photon attenuation off of the optical/UV background
\citep{Fermi_EBL2012,Domi_etal:13}. 

There does not appear to be any evolution of $k_{mM}$ with source
flux, and thus dimmer sources appear to exhibit the same shape
correction as brighter objects.  Nevertheless, by its definition, the
1FHL includes only sources that have been detected above 10~GeV, and
thus it is not immediately obvious that for sources below the
detection threshold that this remains true.  Figure
\ref{fig:kcorr_2FGL} shows the same analysis for all power-law objects
within the 2FGL (the class of sources we will employ henceforth, by
virtue of the applicability of the power-law band correction).  In
this case, there are many upper limits associated with non-detections
(denoted by the blue triangles in Figure \ref{fig:kcorr_2FGL}).  These
form a plateau at higher measured fluxes than our band-corrected
value.  Thus, there is at present no evidence for a flux-dependent
softening at or just below the detection threshold.

It is important at this stage to note that $\left<k\right>$ is {\em
  not} what matters for the definition of the $C_P$.  Rather, since it
is the square of the flux that enters, it is the square of the
spectral shape correction, and thus since there is no apparent
correlation between the spectral shape and flux, $\left< k^2 \right>$.
By definition, this is larger than $\left< k\right>^2$, and even for
the 30~GeV--100~GeV band greater than unity.  That is, the contributions
from the highest flux sources will dominate the anisotropy spectrum,
resulting in a moderation in the impact of a decreasing mean flux
when the scatter in the distribution rises.  Thus, below 100~GeV, spectral
softening typically makes $\lesssim10\%$ correction to the
band-corrected fluxes of individual sources and $\lesssim15\%$
enhancement in their contributions to the anisotropy. 

In the following, we present anisotropy power spectra for different
energy bands that have been ``band corrected'' in two steps.
(1) Considering only sources that are well fit by a power-law spectrum,
we make a gross band correction for each object and compute the
resulting anisotropy power spectra $C_P$. (2) We correct each band
power of $C_P$ with our empirically found correction factor $\left<
k^2 \right>=1.13$ for residual spectral curvature (see
Appendix~\ref{sec:K-shape-corr}), assuming this correction factor to
be independent of source flux.

\section{Direct Estimates of the EGRB Anisotropy} \label{sec:direct}

The direct methods are characterized by how the $w_j$ in Equation
(\ref{eq:dNdFdGamma}) are chosen: set either to zero or unity
depending on whether or not the source in question would have been
masked out via the procedure followed in \citet{Fermi_aniso}.  

\begin{figure}
\begin{center}
\includegraphics[width=\columnwidth]{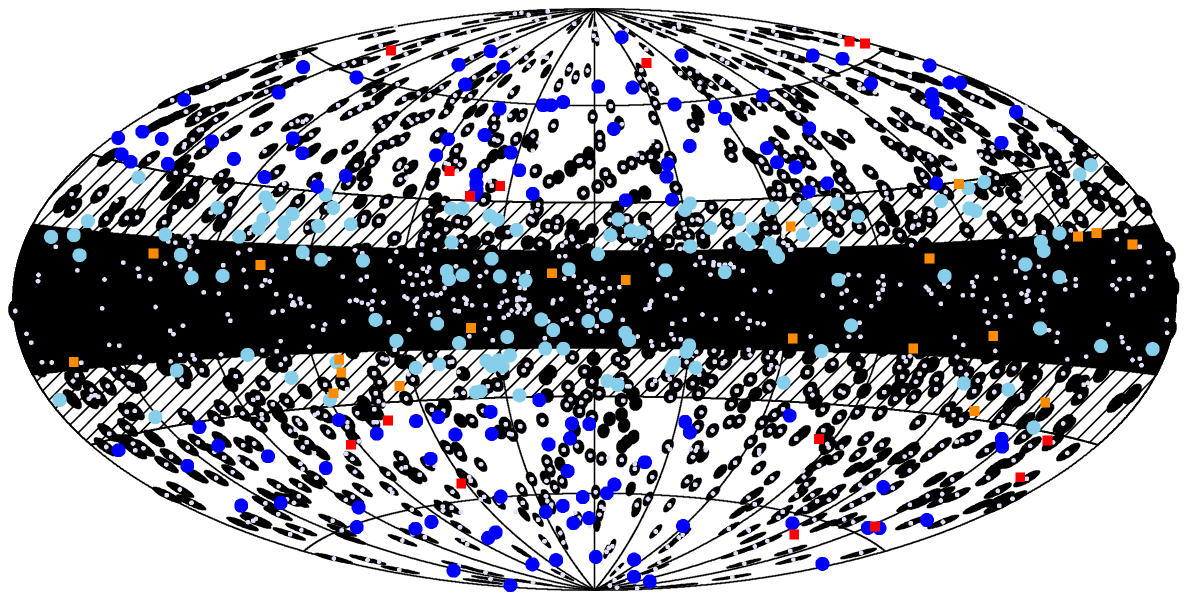}
\end{center}
\caption{Aitoff projection of the 1FGL sky mask employed (black
  region).  The low Galactic-latitude region ($|b|\le30^\circ$) is
  shown by the hatched area and the Galactic plane ($|b|\le15^\circ$)
  is shown in back.  Masked 2FGL sources are shown by the
  gray dots.  Unmasked 2FGL sources detected above 10~GeV are denoted
  by red (orange) squares at high (low) latitudes.  Unmasked 2FGL
  sources with only upper limits in the 2FGL above 10~GeV are denoted
  by blue (light blue) circles at high (low) latitudes.
}\label{fig:mask}
\end{figure}

We reproduce the mask employed by \citet{Fermi_aniso} by first
excluding sources within $2^\circ$ circular regions centered on all
sources reported in the 1FGL and then by applying a Galactic latitude
cut, e.g., $|b|>30^\circ$.  The resulting mask is shown in Figure \ref{fig:mask}.  
Unlike \citet{Fermi_aniso}, since we are estimating the contribution
from resolved point sources directly, we are not limited by the
contaminating diffuse Galactic emission component.  Thus, we are able
to consider alternative Galactic latitude cuts, providing some measure
of the role cosmic variance plays.  Upon comparing the point-source
populations at various potential Galactic latitude cuts, we find a 
strong similarity in the point-source populations for $|b|>15^\circ$
(see Appendix \ref{sec:CV}).  Hence, in what follows, we show both the
$|b|>30^\circ$ mask employed 
by \citet{Fermi_aniso} as well as results associated with a somewhat
conservative, but nonetheless more complete, $|b|>20^\circ$ cut.

Following the implementation of the mask, the resulting value of $C_P$
is then obtained directly from Equation (\ref{eq:CP}), giving
\begin{equation}
  \label{eq:direct}
  \Delta C_P = \frac{1}{\Omega_{\rmn{sky}}}\,\sum_j \F_j^2,
\end{equation}
where the sum extends over all unmasked sources and
$\Omega_{\rmn{sky}} =4\pi f_{\rmn{sky}}$. Values of the unmasked sky
fraction, $f_{\rmn{sky}}$, are obtained explicitly via a Monte Carlo
integration of the respective sky masks (see Appendix~\ref{sec:CV}).  Note that this is
necessarily a lower limit; it both fails to include any potential
diffuse component and the contributions from sources below the
2FGL/1FHL detection thresholds.  For this reason, we denote the EGRB
anisotropy spectrum estimates obtained here by $\Delta C_P$, with the
understanding that they can represent only a fraction of the total
values.

Here we describe a variety of limits of the form described above,
distinguished by the point-source catalog (2FGL and 1FHL) and estimate
of the fluxes, $\F_j$, used.

\begin{figure}[t!]
\begin{center}
\includegraphics[width=0.9\columnwidth]{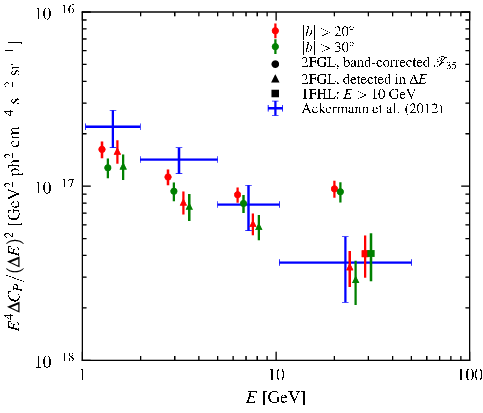}
\end{center}
\caption{ Contribution to the EGRB anisotropy $\Delta C_P$ by unmasked
  point sources employing two different latitude cuts of
  $|b|>20^\circ$ (red) and $|b|>30^\circ$ (green). We compare the
  reported measurements of $C_P$ \citep[blue,][]{Fermi_aniso} to the
  contribution of individual unmasked 2FGL sources, which are detected
  in the respective energy bins (triangles) and the unmasked 1FHL
  sources above 10 GeV (squares). The contribution of all unmasked
  2FGL sources (detected above 1 GeV) that are band corrected to the
  respective energy band (circles) already overproduce the \Fermi
  measurement by $3.2\sigma$ ($3.4\sigma$) in the highest energy band
  for $|b|>30^\circ$ ($|b|>20^\circ$). The error bars indicate the
  cosmic variance and include the propagation of flux
  uncertainties. The points have been shifted horizontally within
  their respective energy bin centers for clarity.}
\label{fig:direct_main}
\end{figure}

\subsection{$C_P$ from Unmasked Sources with Detections in each Energy Band}

In our most conservative approach, we {\em only} use sources with detections
in the respective energy bands. Since the energy bands reported in the
catalogs do not correspond to the energy band of the reported $C_P$,
the flux computation requires some spectral correction, which
depends upon the spectral shape. Here, we only use those
point sources that are well fit by a power-law spectrum with spectral
index $\Gamma$ (and are classified as such in the catalogs), making the
standard band correction for each source individually, using Equation
(\ref{eq:Kdef}) to relate the fluxes in the target band ($E_{m,M}$)
and the lower and upper energy limits of the bands reported in the
catalogs ($E_{x,y}$).  We apply the following band corrections to 2FGL
sources.
\begin{equation}
  \label{eq:Kcorr1}
  \nonumber
  \begin{array}{llll}
    1-3    \mbox{ GeV} &\to& 1.04-1.99 \mbox{ GeV,}  & \mbox{for 1-3 GeV sources}, \\
    1-10   \mbox{ GeV} &\to& 1.99-5    \mbox{ GeV,}  & \mbox{for 3-10 GeV sources}, \\
    3-10   \mbox{ GeV} &\to& 5-10.4    \mbox{ GeV,}  & \mbox{for 3-10 GeV sources}, \\
    10-100 \mbox{ GeV} &\to& 10.4-50   \mbox{ GeV,}  & \mbox{for 10-100 GeV sources}. \\
  \end{array}
\end{equation}
In the case of the 1FHL catalog, we band correct the $10$--$30$~GeV fluxes
to $10.4$--$50$~GeV. As a result, we find 15 (21) unmasked sources for
$|b|>30^\circ$ in the highest-energy band of the 2FGL catalog
(respectively the $10$--$30$~GeV band of the 1FHL catalog) that
contribute to $\Delta C_P$.  As shown in Figure~\ref{fig:direct_main}, in the
case of the 1FHL catalog, these 21 sources already explain the
measured anisotropy in the high-energy band. 

Note that this can only be a lower limit to the true anisotropy since
in addition to the restrictions mentioned at the end of the preceding
section, here it is further assumed that sources detected at lower
energies in the 2FGL---which are not significantly detected at the
high-energy band due to the decreasing \Fermi-LAT sensitivity
there---would not emit any flux at $E>10$~GeV. 
This latter assumption is even more severe than assuming that the 2FGL
is complete.

\subsection{$C_P$ from all Unmasked 2FGL Sources}

Within the 2FGL, we find 139 unmasked power-law sources with
$|b|>30^\circ$, many more than the 15 that have explicit detections
above 10~GeV.  While possible, we consider it extraordinarily
unnatural to believe that sources lying just below the high-energy
\Fermi~detection threshold are completely devoid of high-energy
gamma-ray emission, if only because these sources appear in all other
ways to form a continuous distribution with those that have been
detected.  Thus, in principle, all of these should contribute to the
$\Delta C_P$.  Here, we assess the point-source contribution to the EGRB
anisotropy spectrum using fluxes band corrected from the 1--100~GeV
band ($\F_{35}$) for all sources in the 2FGL.  Note that this still
assumes that the 2FGL is complete, and thus remains a {\em lower}
limit on the expected $\Delta C_P$.

In Figure~\ref{fig:direct_main}, we show the contribution of all
unmasked 2FGL power-law sources (i.e., detected above 1 GeV) that are
band corrected to the respective energy band. At the highest-energy
band, the inferred $\Delta C_P$ values {\rm exceed} the \Fermi-LAT
measurement by $3.2\sigma$ ($3.4\sigma$) for $|b|>30^\circ$
($|b|>20^\circ$). That is using only the unmasked 1FHL and 2FGL
sources, the lower limit upon the anisotropy either fully accounts
for, or significantly exceeds the reported values, respectively!

\section{Statistical Estimates of the EGRB Anisotropy} \label{sec:statistical}

The preceding direct estimates employed the realization of gamma-ray
point sources present in the unmasked 2FGL/1FHL.  However, both to
reduce cosmic variance, which appears to produce an anomalously low
anisotropy signal near latitude cuts of $30^\circ$ (see Appendix
\ref{sec:CV}), and to provide a better comparison to theoretical
models of the gamma-ray point-source population, we also provide a set
of statistical estimates of the EGRB anisotropy.
These employ the entire 2FGL as a statistical representation of the
point-source population.  Key to this is the assumption of the
isotropy (i.e., the masked and unmasked 2FGL sources are statistically
similar) and a characterization of the 1FGL detection threshold,
corresponding to the computation of the $w_j$ in Equation
(\ref{eq:CP}).  We treat both of these here before describing the
corresponding estimate for the EGRB anisotropy spectrum.

\subsection{Isotropy of the 2FGL}
Clustering of the 2FGL sources would presumably decrease their
representation in the sample after application of the mask.  However,
evidence for this isotropy may be found in the investigation of mask
dependence in \citet{Fermi_aniso}\footnote{We do not appeal directly
  to the EGRB power spectrum since to do so we would beg
  the question, having assumed that the background is due nearly
  exclusively to point sources.  Nonetheless, that we do find this to
  be the case, in retrospect, the EGRB power spectrum provides powerful
  confirmation of this assumption.}.  Assuming Poisson statistics,
with 154 more high-latitude sources ($|b|\ge30^\circ$) found in the
2FGL than in the 1FGL, the anticipated ratio of unmasked sky fractions
is 
\begin{equation}
\frac{f_{\rm sky}^{\rm 2FGL}}{f_{\rm sky}^{\rm 1FGL}}
\simeq
e^{-(\Omega_{\rm psm}/\Omega_b)\Delta N}
=
0.910\,,
\end{equation}
where $\Omega_{\rm psm}=\pi(2^\circ)^2=3.83\times10^{-3}~{\rm sr}$ is
the solid angle of the point-source mask and 
$\Omega_b(|b|\ge30^\circ)=2\pi$ is the solid angle residing
above the Galactic latitude mask.  This agrees nearly exactly with 
the stated ratio in \citet{Fermi_aniso}, $0.295/0.325 = 0.907$, and
our own estimate (see Appendix \ref{sec:CV}), $0.310/0.338=0.917$,
providing some confidence that clustering may be neglected.  A more
complete comparison of source properties with various latitude cuts is
presented in Appendix \ref{sec:CV}, where it was found that at
Galactic latitudes above $15^\circ$ the source populations are
statistically similar (i.e., have similar flux and spectral
properties) and have nearly identical numbers of point sources per
unit solid angle at different Galactic latitudes (i.e., once the
Galactic component no longer contributes significantly, the number of
objects per square degree is fixed within the expected Poisson
fluctuations).

\subsection{The 1FGL Detection Threshold}\label{sec:threshold}

\begin{figure}
\begin{center}
\includegraphics[width=0.8\columnwidth]{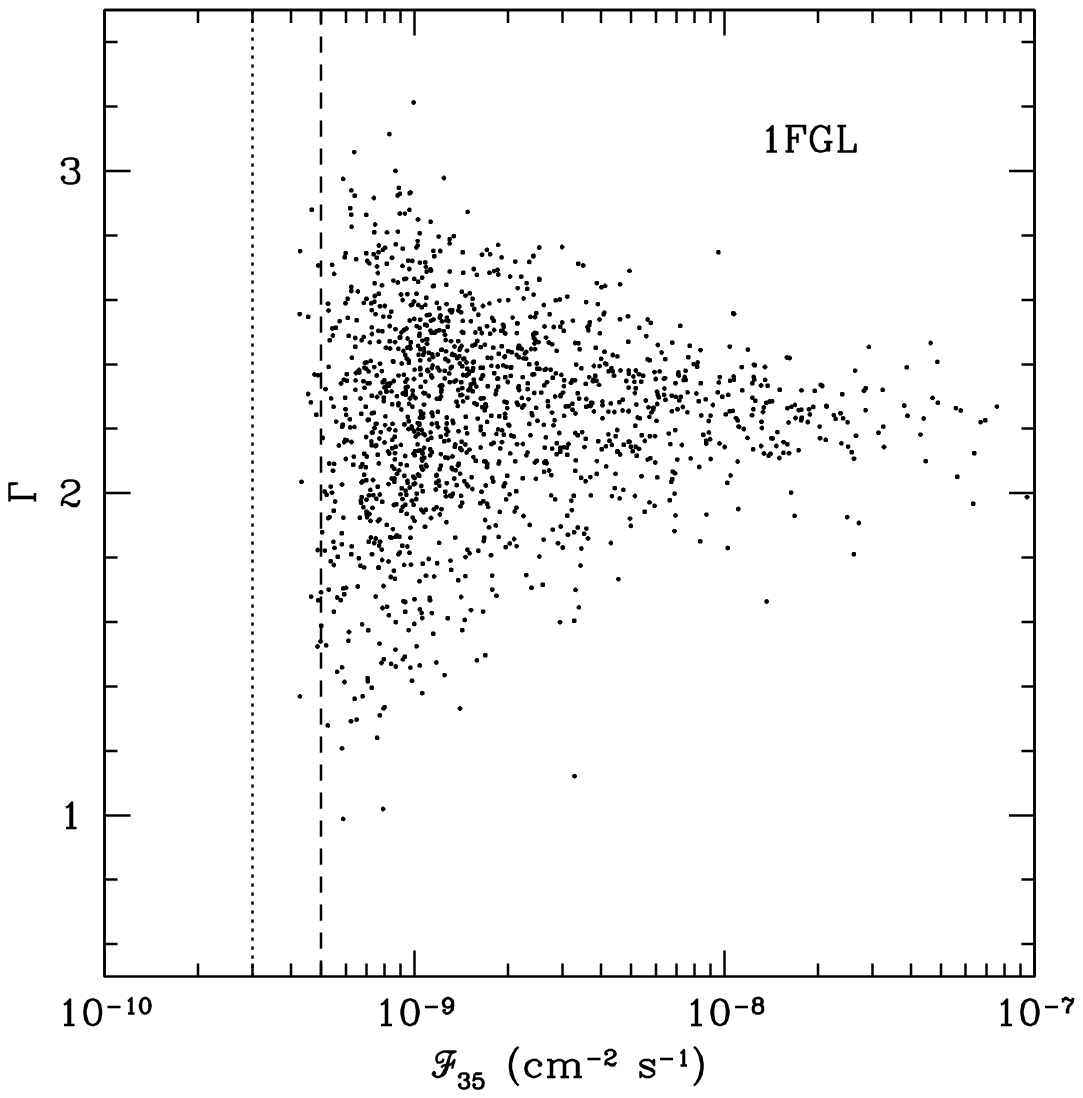}
\includegraphics[width=0.8\columnwidth]{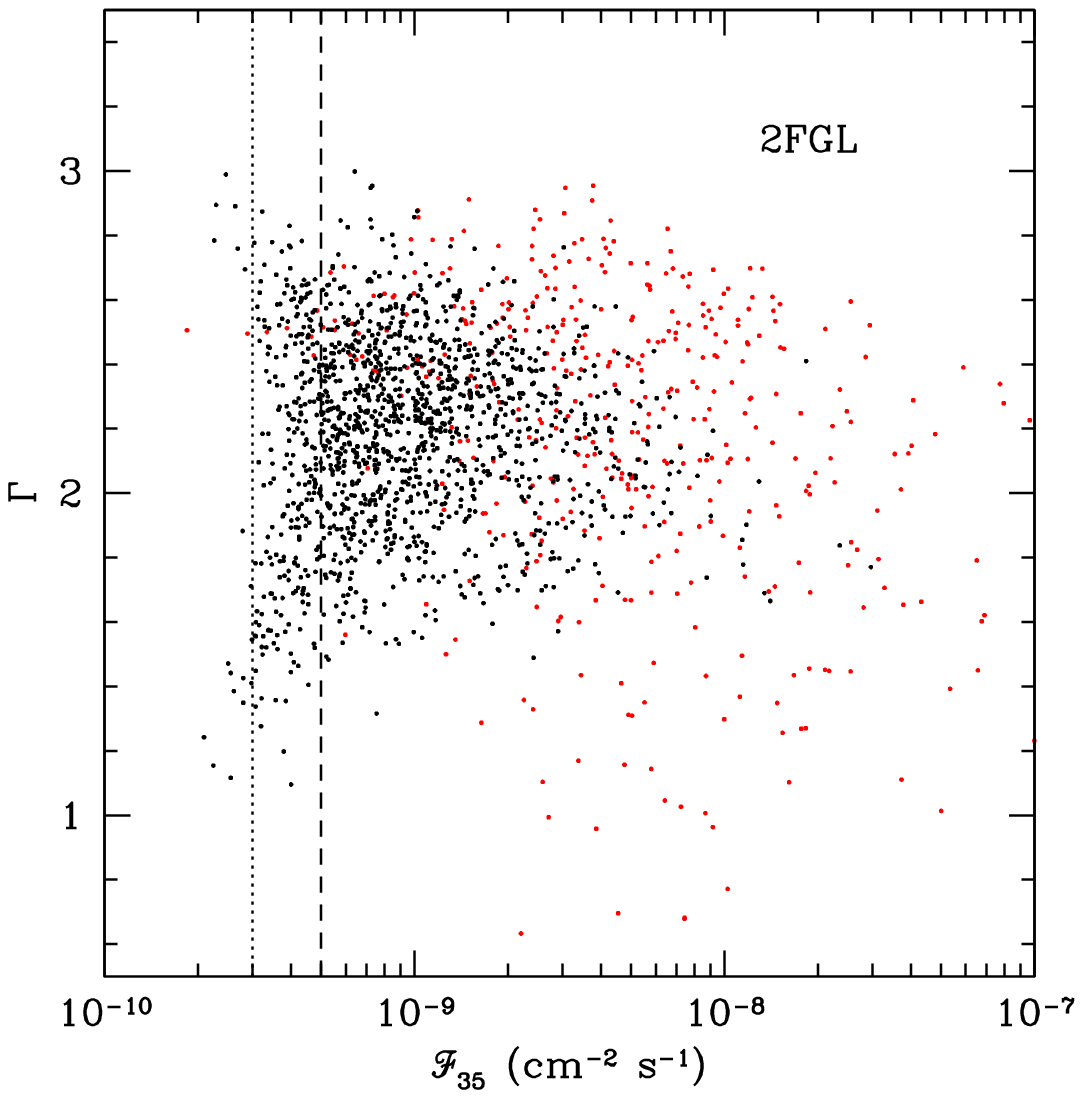}
\end{center}
\caption{Photon spectral index vs flux for the 1FGL (top) and 2FGL
  (bottom). The vertical dotted lines show very roughly the flux limits
  of the 1FGL (dashed) and 2FGL (dotted).  Red points in the
  latter indicate sources with spectra that are better fit with
  non-power-law models.  These are excluded in the estimate of the
  EGRB anisotropy limits.}\label{fig:FGL_GvF} 
\end{figure}

In the direct measures, the 1FGL detection threshold entered implicitly
through the mask.  Here, we use the observed 1FGL source distribution
to estimate the detection threshold explicitly and therefore assign
explicit values to the $w_j$.  In practice, this  is consistent with a
condition solely upon the photon flux from 1 to 100~GeV ($\F_{35}$).
Importantly, there is no systematic correlation between $\F_{35}$ 
and the photon spectral index, $\Gamma$ \citep[see the top panel
  of Figure \ref{fig:FGL_GvF} and Figure 1 of][]{Cuoco}. This is very
different from the photon flux from 100~MeV to 100~GeV ($\F_{25}$),
which strongly correlates with the photon spectral index for small
fluxes due to the strong energy dependence of the point-spread
function of \Fermi-LAT below 1 GeV \citep[see Appendix A
of][]{1FGL,Cuoco}. Hence, the intrinsic flux distribution function can
either be directly studied above 1 GeV or by adopting a non-parametric
method, which allows for reconstruction of it from the observationally biased
flux distribution \citep{SPA12}. In any case, the detection threshold
is not, however, consistent with a simple cutoff, and in the interest
of completeness, we characterize the threshold function here.

\begin{figure}
\begin{center}
\includegraphics[width=0.9\columnwidth]{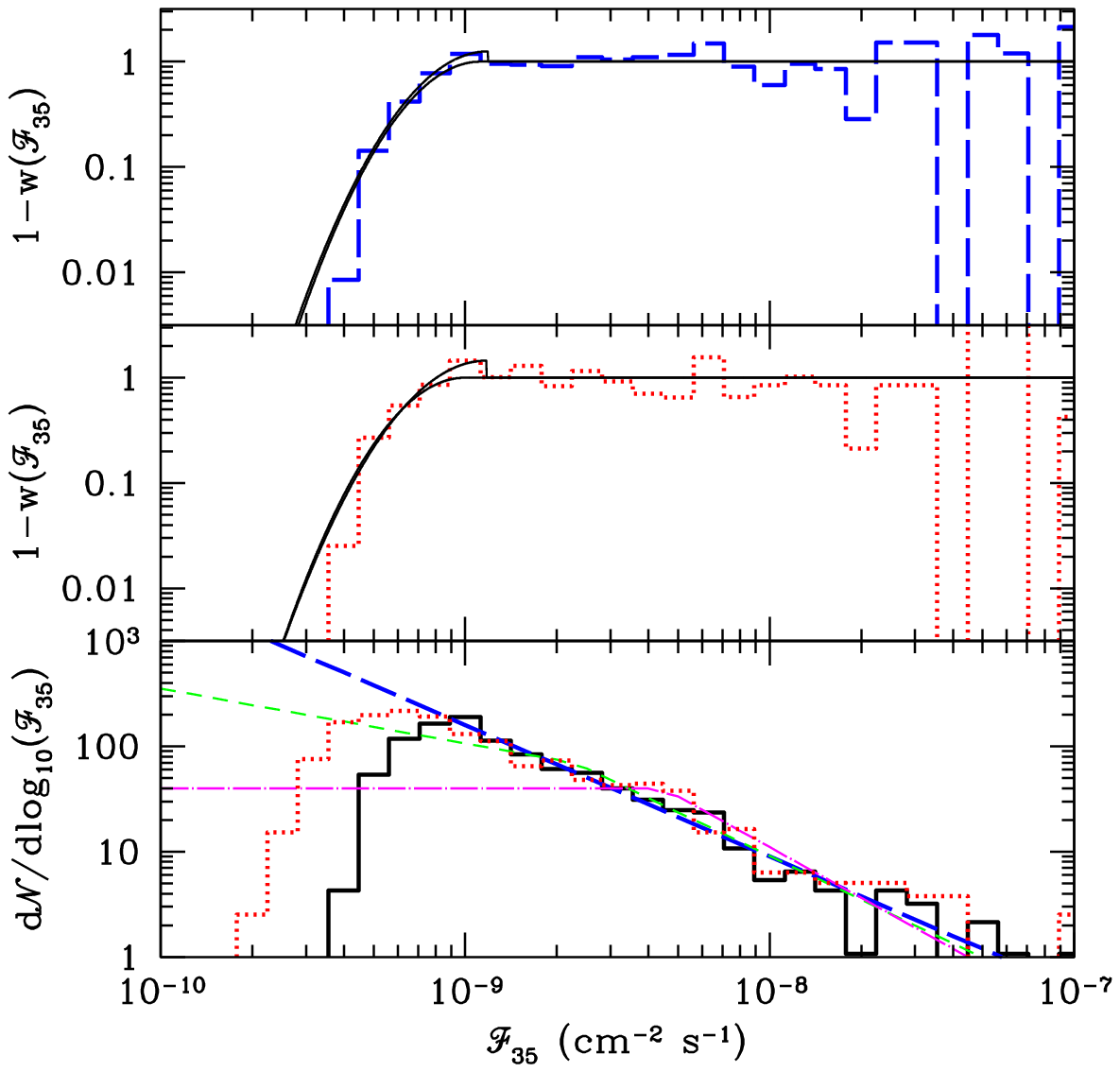}
\end{center}
\caption{ Bottom: distribution of 1FGL sources with 1~GeV--100~GeV
  flux (black), in comparison to two models for the true source
  distribution.  At high fluxes, the distribution is well fit by a
  single power law (dashed blue line), which provides an upper limit
  upon the inferred source population at low fluxes.  An absolute
  lower limit comes from the distribution of sources in the 2FGL,
  normalized to be consistent with the 1FGL for 
  $\F_{35}>10^{-9}~{\rm ph~cm^{-2}~s^{-1}}$ (dotted red line).  
  In addition, for comparison we also show the flux distributions
  employed in \citet[][dashed green line]{Cuoco}, and
  \citet[][dash-dot magenta line]{Harding:2012}.  Note that these are
  meant to reproduce the subpopulation of Fermi point sources
  corresponding to blazars.  Middle and top: inferred detection
  efficiencies, $1-w(\F_{35})$, for the two approximations of the true
  source distribution shown in the bottom panel.  In both cases, we
  show a best-fit non-negative $w(\F_{35})$ of the form described in
  the text and a version that allows $w(\F_{35})$ to extend below
  zero (detection efficiencies greater than unity) to describe the
  potential impact of Eddington bias upon the low-flux tail of the
  1FGL source distribution.  These bound the plausible range of
  detection efficiencies.}\label{fig:cut}
\end{figure}

It is not possible to estimate the detection threshold from the 1FGL
alone; as a result, some assumption must be made regarding the
undetected source population.
The flux distribution of the 1FGL is well approximated by a power law
at high fluxes:
\begin{equation}
\frac{d\N}{d\log_{10}\F_{35}}
\simeq
1.8\times10^3\left(\frac{\F_{35}}{10^{-9}~{\rm ph~cm^{-2}~s^{-1}}}\right)^{-1.25}\,,
\label{eq:dNdlogF}
\end{equation}
(see Figure \ref{fig:cut}; note the above is corrected for sky
fraction), providing a plausible upper bound upon the number of
undetected sources.\footnote{The power law listed here differs from
  those reported in \citet{Fermi_lNlS} as a result of our using the
  1~GeV--100~GeV fluxes.}
Also shown in Figures \ref{fig:FGL_GvF} and \ref{fig:cut} is the 2FGL
population.  As anticipated, it extends to marginally lower fluxes,
resolving a portion of the gamma-ray background unresolved by 1FGL,
roughly consistent with the decrease expected from the increase in
exposure time.  At high fluxes, the normalized 1FGL and 2FGL 
are consistent with being drawn from the same population, i.e., above
the 1FGL detection threshold, the source population flux distributions
are statistically similar.
Making the conservative
assumption that the 2FGL is complete, it provides an absolute lower
bound upon the number of undetected sources.  More complicated source
populations discussed in the literature fall between these two limits
\citep[see, e.g.,][]{Fermi_lNlS,SPA12}. 

For comparison, the flux distributions employed previously by
\citet{Cuoco} and \citet{Harding:2012} are also shown in 
Figure \ref{fig:cut}.  We note that the subpopulation of gamma-ray
sources comprised of blazars were being modeled in both cases.
Nonetheless, it is clear that neither accurately reproduces the 1FGL
or 2FGL source distribution, both exhibiting a break above the 1FGL
flux threshold and substantially underestimating the number of
gamma-ray sources immediately below it.  This is a consequence of
having been originally constructed in a different flux
band, $\F_{25}$, for which the detection threshold exhibits a strong
correlation with the source spectrum below 
$10^{-9}~{\rm ph~cm^{-2}~s^{-1}}$.  Near this flux the observed flux 
distribution exhibits a break \citep{Fermi_lNlS}, though both its
proximity to the flux at which detection threshold becomes independent
of the photon spectral index and its absence in the $\F_{35}$ flux
distribution suggest that it is strongly impacted by the particular
detection efficiency adopted.  This is supported by the lack of any
such feature in the 1FHL flux distribution after a careful debiasing
near the flux threshold was performed 
\citep[see, e.g., Figure 32 of][]{1FHL}.  Despite the critical
differences at low fluxes, all estimates are similar at sufficiently
high fluxes, where the impact of the detection efficiency is minimal.

\begin{deluxetable}{llccc}
\tablecaption{Weight function Parameters for the 1FGL\label{tab:wfits}}
\tablehead{ Threshold Model & Source Distribution & $\F_{\rm max}$\tablenotemark{a} & $m$ & $y$ }
\startdata
2FGL & 2FGL & 1.00 & 7 & 0\\
2FGL-EB & 2FGL w/Eddington bias & 1.18 & 6 & 0.176\\
PL & Power law & 1.12 & 7 & 0\\
PL-EB & Power law w/Eddington bias & 1.19 & 6.5 & 0.097
\enddata
\tablenotetext{a}{Fluxes are in units of $(10^{-9} {\rm ph~cm^{-2}s^{-1}})$.}
\end{deluxetable}

In practice, due to its abrupt nature, the detection threshold is only
weakly sensitive to the form of the low-flux extension assumed.  In
both cases, it is well modeled by a log-normal cutoff.  That is, the
probability of {\em non-detection} is
\begin{equation}
w(\F_{35}) \simeq 
\begin{cases}
1-10^{-m [\log_{10}(\F_{35}/\F_{\rm max})]^2 + y} & \F_{35}<\F_{\rm max}\\
0 & \text{otherwise}\,,
\end{cases}
\label{eq:w1FGL}
\end{equation}
where the values of the various constants are listed in Table \ref{tab:wfits}.
The threshold flux, $\F_{\rm max}$, is consistently near 
$10^{-9}~{\rm ph~cm^{-2}s^{-1}}$, and the steepness parameter, $m$, is
near $6.5$.  While this consistently overpredicts the detection
efficiency at very low fluxes, it provides a good approximation near
the threshold and therefore at the fluxes that dominate the
contribution to the EGRB anisotropy. The
characteristic value of $5\times10^{-10}~{\rm cm^{-2}s^{-1}}$ employed
by \citet{Cuoco} produces a similar threshold, if not slightly lower.

Some care should be taken in interpreting the distribution of sources
near the detection threshold due to the potential for a significant
Eddington bias \citep{Eddi:1913,Eddi:1940}.  That this is present in the 1FGL is clear from the
fact that the number of 1FGL sources right near the threshold 
{\em exceeds} that in the presumably more complete 2FGL.  Deconvolving
the Eddington bias is non-trivial and depends on both the
instrumental and intrinsic fluctuations in the measured source
fluxes.  Here, we take a simplified, if somewhat unphysical approach,
of allowing the detection efficiency to exceed unity immediately at
the threshold, describing the detection of false positives
corresponding to lower-flux sources that are temporally above the
threshold due to statistical and intrinsic fluctuations \citep[a
  similar approach is adopted in][see Figure 30 and the surrounding
  discussion]{1FHL}. This corresponds to a negative $w(\F_{35})$, and
{\em lowers} the EGRB anisotropy. 

Many bright sources, well above the putative 1FGL flux limit, appear
in the 2FGL but not the 1FGL.  Unlike the high-flux behavior, this is
presumably due to variability, i.e., sources that were bright only
following the initial nine months included in the 1FGL.  Similarly, in
the 2FGL, many 1FGL sources lie below the approximate 1FGL flux
threshold, likely due to Eddington bias (both due to intrinsic
variability and statistical fluctuations).

\subsection{Estimating $C_P$ from the 2FGL} \label{sec:2FGL} We
supplement the weights obtained above with a Galactic latitude cut of
$|b|>30^\circ$ to more directly represent the population of relevance
to the results reported in \citet{Fermi_aniso}.  As previously
mentioned, this guarantees that any Galactic point-source contribution
has been eliminated.  Inserting these conditions into Equation
(\ref{eq:CP}), i.e., only accounting for objects with $|b|>30^\circ$
in the summation and employing the weights in Equation
\ref{eq:w1FGL}, then provides an estimate for the $\Delta C_P$.

Note that in this case the source distribution is expressed in terms
of $\F_{35}$ and not the flux over the energy range for which $C_P$
is being constructed, $\F$.  This is necessary since our detection
threshold is specified in terms $\F_{35}$; fluxes over other energy
ranges exhibit substantial spectral-index-induced biases.  That is, we
specify
\begin{equation}
\frac{d\N}{d\F_{\rm 35} d\Gamma} \simeq \sum_j \frac{w_j}{\Omega_{\rm sky}}
\delta(\F_{35}-\F_{35,j}) \delta(\Gamma-\Gamma_j)\,,\label{eq:dNdFdGamma2}
\end{equation}
where $w_j=w(\F_{35,j})$.

In accordance with the particular form of the band correction used and
given the importance of the spectral shape to the reconstruction of
the EGRB anisotropy, we restrict ourselves to the subset of the 2FGL
that is best fit by power-law spectra, comprising roughly 78\% of the
total sample. Since the sources for which non-power-law fits are
statistically favored tend to be bright, this restriction makes an
insignificant difference in our EGRB anisotropy estimates in
practice.

\begin{figure}[t!]
\begin{center}
\includegraphics[width=0.9\columnwidth]{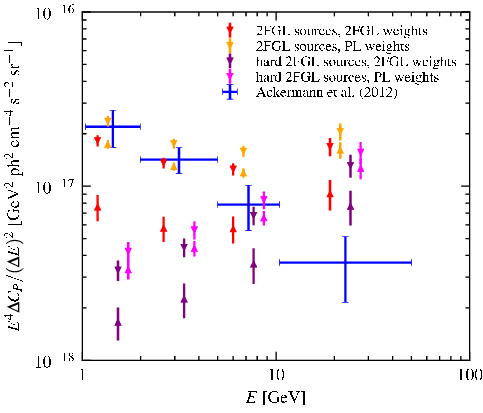}
\end{center}
\caption{Reported EGRB anisotropy spectrum \citep[blue error bars,
  taken from][]{Fermi_aniso} in comparison to the implied
  contributions from high-latitude ($|b|>30^\circ$) 2FGL sources
  (triangles). Downward pointing triangles assume the best-fit
  non-negative $w(\F_{35})$ of the form described in the text, and
  upward pointing triangles allow $w(\F_{35})$ to extend below zero
  (detection efficiencies greater than unity) to describe the
  potential impact of Eddington bias upon the low-flux tail of the
  1FGL source distribution. Orange triangles show the EGRB anisotropy
  spectrum inferred assuming a power-law low-flux extension of the
  1FGL, and the red triangles present the conservative lower limit
  obtained by assuming the 2FGL is complete. For reference, the
  contributions from only hard sources (i.e., $\Gamma\le2$) are also
  shown (purple and magenta). The error bars indicate the cosmic
  variance and include the propagation of flux uncertainties. The
  points are offset horizontally within their respective energy bin
  centers for clarity.}\label{fig:EAlim}
\end{figure}

The estimated contribution to the EGRB anisotropy from the low-flux,
high-latitude sources in the 2FGL is shown by the orange
  triangles in Figure \ref{fig:EAlim}, with the
conservative lower limits arising from assuming the 2FGL is complete
shown by the red triangles. The estimates using
$w_j$ that account for the Eddington bias in the 1FGL (upward
  pointing triangles for each method, respectively) are consistent
with the direct estimates obtained in Section \ref{sec:direct},
yielding some confidence in the statistical approach.  When $w_j$
is constructed assuming a power-law extrapolation of the low-flux source
population are used, the lower limits rise further.

Nevertheless, the anisotropy due to sources resolved by the 2FGL alone
are sufficient to produce the entirety of the EGRB anisotropy from
2--5~GeV and {\em exceed} the reported EGRB anisotropy above 5~GeV by
many $\sigma$.  At high energies, this is due primarily to the
presence of hard gamma-ray sources, i.e., objects for which
$\Gamma\le2$, shown by the magenta and purple
triangles.\footnote{At all energies, however, the 2FGL contribution
to the isotropic component of the EGRB is well below the reported
limits.}  These are dominated by BL Lac objects, and are presumably
responsible for some fraction of the isotropic EGRB component.
However, these are not solely responsible for the anisotropy excess,
playing a subdominant role below 10.4~GeV.

\subsection{Estimating $C_P$ from Power-law Extensions of the 2FGL} \label{sec:statpl}

The assumption made above that only sources that are detected in the
2FGL contribute to the EGRB anisotropy is itself extreme.  While
contributions to the $C_P$ are
heavily biased toward high-flux objects, the range about the flux
threshold for which significant contributions are found extends for at
least a decade.  By comparison, the reduction in the detection
threshold between the 1FGL and 2FGL is roughly 0.2 dex, considerably
smaller.  Thus, the assumption that the 2FGL is complete corresponds
to a dramatic suppression in the number of sources immediately below
the 2FGL detection threshold.  Evidence that this is not the case may
be found in the recently published catalog of hard Fermi sources, the
1FHL, which does not exhibit any notable features at the 2FGL flux
limit.  Hence, the $\Delta C_P$ obtained in the previous sections can
at most represent a modest contribution to the anticipated value.

To provide a reasonable upper limit on the expected contribution to
the EGRB anisotropy due to gamma-ray point sources, we consider the
$C_P$ obtained from a single power-law extrapolation of the 2FGL
population, given in Equation (\ref{eq:dNdlogF}).  This is 
highly uncertain.  To guarantee a finite contribution to the isotropic
EGRB component, a break in the $d\N/d\F_{35}$ relation must 
exist at sufficiently low fluxes.  Similarly, to guarantee a finite
source population, a cutoff must exist as well (though
this is much less well constrained).  To assess the sensitivity to
these putative features, we produce anticipated EGRB anisotropy
spectra for a variety of lower flux cutoffs.

We must supplement the extrapolated $d\N/d\F_{35}$ relation with a
spectral index distribution.  Based on Figure \ref{fig:FGL_GvF}, we
assume the distribution in $\F_{35}$ and $\Gamma$ is separable, i.e.,
we choose
\begin{equation}
\frac{d\N}{d\F_{35}d\Gamma}
=
\frac{d\N}{d\F_{35}}
\frac{e^{-(\Gamma-\bar{\Gamma})^2/2\sigma_\Gamma^2}}{\sqrt{2\pi} \sigma_\Gamma}\,,
\end{equation}
where $\bar{\Gamma}=2.2$ and $\sigma_\Gamma=0.3$, measured from the
2FGL directly.  While the intrinsic photon spectral index distribution
is clearly skewed, the above does an especially good job of
reproducing the hard component, critical for the high-energy EGRB
anisotropy estimates.

\begin{figure}[t!]
\begin{center}
\includegraphics[width=0.9\columnwidth]{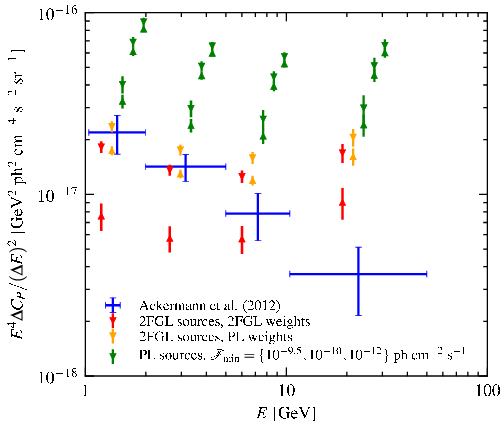}
\end{center}
\caption{Reported EGRB anisotropy spectrum \citep[blue error
    bars, taken from][]{Fermi_aniso} in comparison to that inferred
    from a power-law extrapolation of the 2FGL to lower fluxes.  The
    green triangles show the EGRB spectrum for various low-flux
    cutoffs, $3\times10^{-10}~{\rm cm^{-2}s^{-1}}$, $10^{-10}~{\rm
      cm^{-2}s^{-1}}$, and $10^{-12}~{\rm cm^{-2}s^{-1}}$ (effectively
    zero) from left to right within each energy bin.  Downward
    pointing triangles assume the best-fit non-negative $w(\F_{35})$
    of the form described in the text, and upward pointing triangles
    allow $w(\F_{35})$ to extend below zero (detection efficiencies
    greater than unity) to describe the potential impact of Eddington
    bias upon the low-flux tail of the 1FGL source distribution.  For
    reference, the red triangles show the limits obtained under the
    assumption that the 2FGL is complete (see Figure \ref{fig:EAlim}).
    The error bars indicate the cosmic variance and include the
    propagation of flux uncertainties. The points are offset
    horizontally within their respective energy bin centers for
    clarity.}\label{fig:EApl}
\end{figure}

Figure \ref{fig:EApl} shows the associated EGRB spectrum for a handful
of lower-flux cutoffs.  From this, it is apparent that the $C_P$
receive a substantial contribution from sources that lie below the
2FGL flux threshold.  In the absence of a dramatic suppression of the
source population above $\F_{35}\simeq10^{-10}~{\rm cm^{-2}s^{-1}}$,
these unresolved sources comprises more than half of the anticipated
contribution.

\section{Discussion} \label{sec:discussion}

While the direct estimate of the EGRB anisotropy above 10~GeV from the
1FHL is consistent with the reported values, this requires two
extraordinary assumptions: first, that the 1FHL is complete, and
second, that nearly all sources within the 2FGL that are undetected
above 10~GeV have sufficiently softened spectra that they emit
negligibly there.   Excluding these, our lower limit upon the EGRB
anisotropy spectrum significantly exceeds the values reported by
\citet{Fermi_aniso}.  The reason for this is not immediately clear.
Here, we consider a handful possibilities.

The first is that that the 1FHL and 2FGL are complete and sources
without detections above 10~GeV in the 2FGL and 1FHL really do have
vanishing flux at that energy, i.e., band correcting their low-energy
emission substantially overestimates their true high-energy fluxes.  
However, the former is belied by the absence of any feature near the
flux thresholds of the 1FHL or 2FGL in the flux distributions.  The
latter is argued against by the detection of high-energy emission from
a number of additional sources reported in the 1FHL, though presumably
this may arise from variability.  Perhaps more conclusive is the
evidence that the spectrally corrected $\F_{35}$ provides a good
estimate of the 1FHL measured fluxes (see Appendix
\ref{sec:K-shape-corr}), strongly suggesting that this will be true
for sources falling immediately below the relevant detection
thresholds.  For this reason, we believe this is highly disfavored.

The second is that the deficit in anisotropy is due to a surfeit of
dim sources in the 1FGL resulting from the Eddington bias.  While we have
not made any attempt to debias the 1FGL (which is well beyond the
scope of this paper), this effect alone is incapable of producing the
factor of two reduction necessary to resolve the discrepancy.  

The third is the PSF deconvolution procedure employed by
\citet{Fermi_aniso}.  The most recent PSF estimates are much broader
than those used, implying a corresponding underestimate of the
deconvolved $C_P$.  As shown in Appendix \ref{sec:PSF}, rough
estimates suggest that this can amount to a factor of $\sim2$
correction under reasonable assumptions, though the full impact of the
updated instrument response functions (IRFs) requires a full reproduction of
the analysis in \citet{Fermi_aniso}.  When coupled with the cosmic
variance, this can boost the reported anisotropy to the minimum limits
implied by the 2FGL.  However, it is unable to increase it to the
values suggested by smooth extrapolations of the 2FGL population to
significantly lower fluxes.

Finally, the independence of the multipoles of the power spectrum of
the masked gamma-ray sky fails for a small number ($N\lesssim100$) of
weak sources that dominate the anisotropy power spectrum.  As shown in
Appendix \ref{sec:CPvar}, the covariance between different multipoles
induces a floor on the statistical uncertainty comparable to that
reported at high energies.  In the case of the highest-energy bin, where
the anisotropy is due entirely to the 21 unmasked sources in the 1FHL,
the corresponding uncertainties are expected to be 10\%--50\% larger
than those reported.

\section{Conclusions} \label{sec:conclusions} 
The contribution to the EGRB anisotropy arising from the now resolved
point sources in the 2FGL and 1FHL detected above 5~GeV and lying
outside of the mask employed by \citet[][which removed the galactic
  plane and all 1FGL point sources]{Fermi_aniso} is sufficient to fully
explain the reported values at these energies, where BL Lac objects dominate
the resolved extragalactic gamma-ray sky and the EGRB anisotropy has
proved most constraining.  This would imply that there
cannot be any significant contribution from sources that fall just
below the 2FGL and 1FHL detection thresholds, a conclusion belied
by the absence of any evidence of spectral evolution with source flux.
Thus, the unmasked 1FHL and 2FGL sources are in tension with the
reported values of the anisotropy based on removing the 1FGL point sources.

Estimates of the EGRB anisotropy which employ the full 2FGL,
extrapolating the flux contributions at all energy ranges via an
empirically vetted band correction, significantly exceed the reported
1FGL-based measurements above 5~GeV.  This represents a purely
empirical estimate of the anisotropy associated with the known
point-source population.  In particular, we do not appeal to a parameterized
source distribution, and hence our estimate represents a conservative
lower limit on the EGRB anisotropy.  Given that the 2FGL and the EGRB
anisotropy observations are largely coincident, reconciling the
apparent disparity between the estimates presented here and those
reported in \citet{Fermi_aniso} is difficult.  For estimates that make
reasonable assumptions regarding the high-energy fluxes of 2FGL
sources, the discrepancy above 10.4~GeV ranges from 2.6$\sigma$
(direct) to 5.2$\sigma$ (statistical, no Eddington bias), with the
2FGL contribution alone exceeding that reported in \citet{Fermi_aniso}
by a factor of $2.6$--$4.6$, respectively.

The 2FGL limits are quite robust, independent of many of the
complications that plague estimating the EGRB anisotropy
directly\footnote{Of course, this comes at the price of resolving the
  background, and thus is not a reasonable alternative.}.  In
particular, it is insensitive to the particulars of the low-flux
source population (which can only increase the anisotropy), and the
dominant remaining source of potential uncertainty, cosmic variance,
is small in comparison to the uncertainties on the reported anisotropy
values.  Given the lack of a precipitous decline in the number of
sources immediately below the 1FGL threshold observed in the 2FGL, and
more recently, in the 1FHL below the 2FGL threshold, the observed
discrepancy is unsurprising; the 2FGL overproduces the EGRB
anisotropy for precisely the same reason that models of the gamma-ray
point-source distributions do.

The limited flux range below the 1FGL threshold probed by the 2FGL
implies that extrapolations of the 2FGL population to fluxes below the
2FGL threshold exceed the 2FGL limits by a substantial amount.
Barring a cutoff immediately below the 2FGL limit, our estimates of
the point-source contribution to the EGRB is an underestimate by at
least a factor of two at high energies.  As a consequence of the 2FGL
detection threshold the 2FGL contribution to the EGRB anisotropy
primarily probes the nearby AGN population \citep[see Figure 36
of][]{2LAC}.  That is, unlike the EGRB anisotropy, generally, it does
not provide an independent measure of the high-redshift AGN
distribution.  Thus, it is not surprising that models that reproduce
the characteristics of the 2FGL are consistent with our implied limits
upon the EGRB anisotropy \citep[e.g.,][]{PaperV}.  Similarly, in the
absence of a better understanding of the discrepancy between estimates
of the EGRB anisotropy at high energies, as derived here from considering
2FGL and 1FHL sources versus those in \citet{Fermi_aniso} from the masked
1FGL, it is unclear if they provide an independent constraint on
the high-redshift gamma-ray universe at present.

\begin{appendix}

\section{Energy Band and Spectral Shape Corrections}

Computing the flux within the specific energy bands relevant for
comparison to the measured EGRB anisotropy generally requires some
spectral correction from fluxes measured in bands for which data is
readily available.  This depends on the spectral shape.  We implement
a band correction in two steps:
\begin{enumerate}
\item Considering only sources that are well fit by a power-law
  spectrum, we make a gross band correction for each object.  With
  these, we construct the anisotropy spectrum. 
\item We correct the above $C_P$ for the observed spectral softening
  using a shape correction factor estimated by comparing the
  band-corrected fluxes to measured band-specific fluxes.
\end{enumerate}
We describe each in detail here.

\subsection{Band-correcting $C_P$ using Power-law Spectra}
The first band correction employs Equation (\ref{eq:Kdef}) to
construct the band-specific fluxes from the 1~GeV--100~GeV fluxes,
i.e.,
\begin{equation}
\F_j = K(\Gamma_j) \F_{35,j}
\quad\text{where}\quad
K(\Gamma) \equiv
\frac{E_m^{1-\Gamma}-E_M^{1-\Gamma}}{1-100^{1-\Gamma}}\,,
\end{equation}
where $E_{m,M}$ are the lower and upper energy limits of the band in
GeV, and $\F_{35}$ is our standard flux reference.

Note that while the above provides the band correction for
{\em individual} sources, since their individual spectra differ it is
not equivalent to a single band correction, evaluated at some 
effective spectral index.  At high energies, the population
near the 1FGL detection threshold is dominated by increasingly hard
sources, imparting an energy dependence on the typical spectral index
and generally enhancing the anisotropy 
\citep[see, e.g., Section VI of][]{Cuoco}.

\subsection{Spectral Shape Corrections}
\label{sec:K-shape-corr}

Applying the band correction described above assumes that the
source spectra are well approximated by power laws.  Despite their
characterization as such at lower energies, it is not clear that at
high energies that this remains the case.  In particular, above 100~GeV,
considerable softening is anticipated for sources with redshifts
$\gtrsim0.5$.  Thus, prior to using the band correction, here we assess
its applicability and estimate the relevant spectral corrections.  We
do this by comparing the band-corrected $\F_{35}$ to the fluxes measured
in individual energy bands near those of interest.  

For this purpose, we make use of both the 2FGL and 1FHL catalogs.
Because of its longer duration and focus on sources above 10~GeV,
the 1FHL affords better energy resolution and thus we ultimately make
our correction based upon the subset of objects that appear in both
the 1FHL and 2FGL.  Nevertheless, to assess the impact of potential
biases resulting from the 2FGL detection threshold, we also consider
2FGL sources alone.  To exclude any putative Galactic
component, we impose a cut on Galactic latitude of $|b|>30^\circ$.

\begin{figure}[t!]
\begin{center}
\includegraphics[width=\columnwidth]{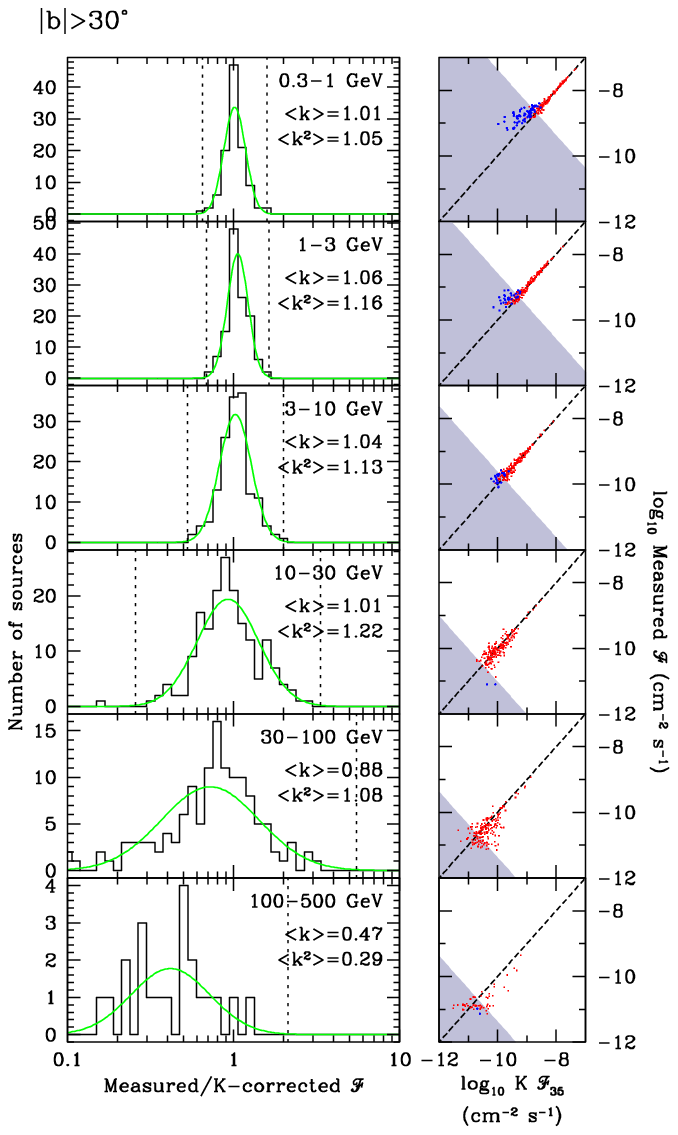}
\end{center}
\caption{Left: distribution of the spectral shape corrections to the
  band-corrected $\F_{35}$ for sources in the 1FHL and 2FGL catalogs for various
  energy bands.  The green line shows a log-normal distribution fit.
  Sources beyond the vertical dashed lines are excluded so as not to
  bias the width toward larger values.  Right: measured vs.~ only
  band-corrected $\F_{35}$ within each energy band.  Red and blue dots denote
  detections and upper limits, respectively.  The dotted line shows
  the equal case, and the shaded area corresponds to the region
  excluded by Equation (\ref{eq:kex}).
}\label{fig:kcorr}
\end{figure}

\begin{figure}[t!]
\begin{center}
\includegraphics[width=\columnwidth]{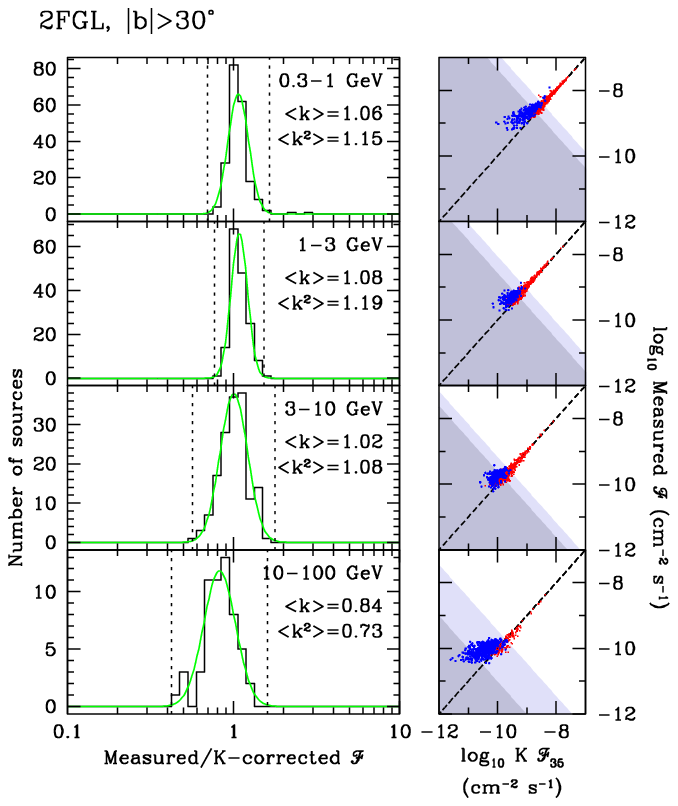}
\end{center}
\caption{Left: distribution of the spectral corrections to the
  band-corrected $\F_{35}$ for all power-law sources in the 2FGL
  catalog for various energy bands.  The green line shows a log-normal
  distribution fit.  Sources beyond the vertical dashed lines are
  excluded so as not to bias the width toward larger values.  Right:
  measured vs.~ only band-corrected $\F_{35}$ within each energy
  band.  Red and blue dots denote detections and upper limits,
  respectively.  The dotted line shows the equal case, and the light
  and dark gray shaded area corresponds to the region excluded by
  Equation (\ref{eq:kex}) for the 2FGL and 1FHL detection
    thresholds, respectively.  Note that all of the upper limits
  (blue points) lie to the left of the trends associated with the
  detections (red points), implying that the detection upper limits
  lie above the band correction estimates.  Thus, there is no evidence
  of a population of misidentified soft sources at high energies.
}\label{fig:kcorra_2FGL}
\end{figure}

In both cases, the measured and band-corrected $\F_{35}$ are clearly
strongly correlated, exhibiting larger scatter at low fluxes,
eventually developing a plateau, typically dominated by upper limits
(see the right-hand panels of Figures \ref{fig:kcorr} and
\ref{fig:kcorra_2FGL}).  Importantly, the detection upper limits
(absent at high energies in Figure \ref{fig:kcorr} due to the
definition of the 1FHL) plateau at measured fluxes {\em above} the
band-corrected values.  This implies that for the power-law sources we
consider, there is no evidence for a flux-dependent evolution in the
spectral shapes, i.e., sources near the detection threshold are
spectrally similar to those above.

To avoid any potential Eddington bias in our estimated shape
correction at low measured fluxes induced by the detection threshold,
we institute a lower flux limit of the form
\begin{equation}
\F_{mM} \times (K \F_{35}) > \F_{\rm min}^2\,,
\label{eq:kex}
\end{equation}
where $\F_{\rm min}$ is an energy-band-dependent flux limit.  The
excluded region is denoted by the shaded regions in the right-hand
panels of Figures \ref{fig:kcorr} and \ref{fig:kcorra_2FGL}, and the
boundary is orthogonal to the expected proportionality relation (and
therefore should not induce any Eddington bias).  Values for the
band-specific $\F_{\rm min}$ adopted here are listed in Table
\ref{tab:kcorr}.

We define the source-specific spectral correction by
\begin{equation}
k \equiv \frac{\F_{mM}}{K \F_{35}}\,,
\end{equation}
which is simply the multiplicative correction to the band correction.
Within each energy band, the distribution of $k$, shown in Figures
\ref{fig:kcorr} and \ref{fig:kcorra_2FGL}, is well approximated by
a log-normal distribution.  The parameters of the distributions are
relatively independent of the particular value of $F_{\rm min}$
employed.  However, among bands, the distributions vary
substantially. In the characterization of the properties of the $k$
distributions, we have excluded outliers, which we define as objects
with spectral corrections located more than three standard deviations
from the mean.  This has the effect of reducing the width of the
distributions and biasing our estimated corrections to the EGRB
anisotropy spectrum toward lower values, making it a conservative
assumption.

At energies less than 30~GeV, the band-corrected fluxes are quite
accurate, with mean spectral corrections, $\left< k \right>$, all of
the order of unity for both the 1FHL and 2FGL samples.\footnote{Note that
  $\log_{10} \left< k\right> = {\mu_k + 0.5\sigma_k^2 \ln10}$ for a
  log-normal distribution with $\mu_k\equiv\left<\log_{10} k\right>$ 
  and $\sigma_k^2 = \left<(\log_{10}k - \mu_k)^2\right>$.  It is for
  this reason that $\left< k \right>$ typically exceeds the mode in
  the distributions in Figures \ref{fig:kcorr} and
  \ref{fig:kcorra_2FGL}.}  At high energies, $\left< k \right>$ is
less than unity, indicating the anticipated softening of the spectra.
Moreover, there is some tentative evidence that this softening becomes
more severe at higher energies.

However, since it is $\F^2$ that enters into the estimate of $C_P$
employed here, of more importance is the mean-{\em square} spectral
correction, i.e., $\left< k^2 \right>$.  This is impacted not only by
the movement in the centroid of the $k$ distribution but also by its
width.  Because of this, $\left< k^2 \right>$ is larger than unity for
all but the highest-energy bin and, by definition, is larger than
$\left<k\right>^2$.  For the 30--100~GeV bin this is despite
$\left<k\right>\simeq0.88$.  The estimates of $\left< k^2 \right>$
vary by about 1.13.

Repeating the above analysis with various other latitude cuts (e.g.,
$|b|>20^\circ$, $|b|>40^\circ$) yields nearly identical results,
implying that the spectral corrections are indeed intrinsic to the
extragalactic source population and are not associated with any
contaminating Galactic subpopulation.  Therefore, assuming that the
distribution in $k$ is uncorrelated with $\Gamma$ or $\F_{35}$, we
adopt a uniform spectral correction of 1.13 below 100~GeV,
corresponding to the average value across bins at these energies.

\begin{deluxetable}{ccc}[th!]
\tablecaption{Kolmogorov-Smirnov Comparison of Latitude Cuts\label{tab:bKS}}
\tablehead{
$b_{\rm min}$
&
$P_{\rm KS}^{\F_{35}}$
&
$P_{\rm KS}^\Gamma$
}
\startdata
$0^\circ$ & $9.2\times10^{-6}$ & 0.13 \\ 
$5^\circ$ & $3.2\times10^{-4}$ & 0.43 \\ 
$10^\circ$ & $9.7\times10^{-3}$ & 0.65 \\ 
$15^\circ$ & 0.15 & 0.70 \\ 
$20^\circ$ & 0.60 & 0.74 \\ 
$25^\circ$ & 0.97 & 1.0 \\ 
$30^\circ$ & 1.0 & 1.0 \\ 
$35^\circ$ & 0.96 & 1.0 \\ 
$40^\circ$ & 0.30 & 0.98 \\ 
$45^\circ$ & 0.23 & 0.48 
\enddata
\end{deluxetable}

\begin{figure*}
\begin{center}
\includegraphics[width=0.9\columnwidth]{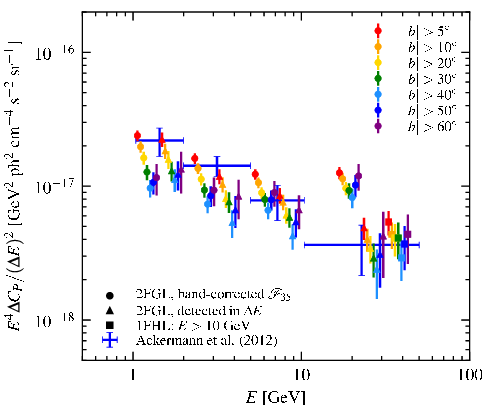}\hspace{2em}
\includegraphics[width=0.9\columnwidth]{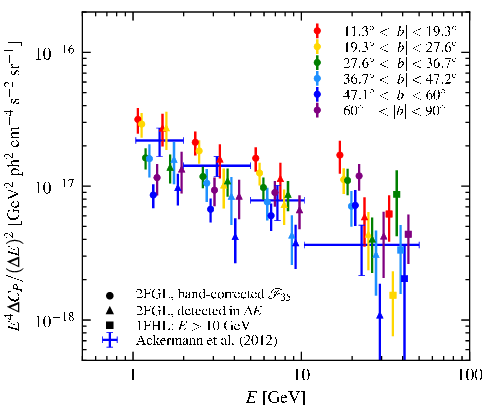}
\end{center}
\caption{Contribution to the EGRB anisotropy $\Delta C_P$ by unmasked
  point sources. We contrast the cumulative (left) and differential
  (right) contribution of galactic latitude (color coded for different
  cuts).  We compare the reported measurements of $C_P$
  \citep[blue,][]{Fermi_aniso} to the contribution of individual
  unmasked 2FGL sources, which are detected in the respective energy
  bins (triangles) and the unmasked 1FHL sources above 10 GeV
  (squares). We show the contribution of all unmasked 2FGL sources
  (detected above 1 GeV) that are band corrected to the respective
  energy band (circles). The error bars account for cosmic variance,
  and the data points have been shifted horizontally within their
  respective energy bin centers for clarity.}\label{fig:direct_app}
\end{figure*}

\section{$C_P$ Limit Estimate Uncertainties}

\subsection{Flux Reconstruction Errors on $C_P$}
\label{sec:error_prop}
Uncertainty in the intrinsic 1~GeV--100~GeV fluxes and the spectral
indexes produce a corresponding uncertainty in the band-corrected
fluxes in the normal way and thus the estimated anisotropy.  Here, we
collect the relevant error propagation formula and some comments on
its magnitude.  Because of its weak energy dependence, we will neglect
the impact of the spectral shape correction, focusing entirely on
the impact of the power-law band correction.

Assuming no correlation in the errors on $\F_{35}$ and $\Gamma$, we
have an uncertainty in the power law band corrected flux of
\begin{equation}
\frac{\sigma_{\F_j}^2}{\F_j^2} = 
  \frac{\sigma_{\F}^2}{\F_{35}^2}
  + 
  \frac{\sigma_K^2}{K^2}\,,
\end{equation}
where
\begin{equation}
\label{eq:sigmaK}
\frac{\sigma_K}{K}
=
\left|\frac{\ln(E_m) E_m^{1-\Gamma} - \ln(E_M) E_M^{1-\Gamma}}{E_m^{1-\Gamma}-E_M^{1-\Gamma}}
-
\frac{\ln(1) - \ln(100) 100^{1-\Gamma}}{1-100^{1-\Gamma}}
\right| \sigma_\Gamma\,,
\end{equation}
in which $E_{m,M}$ are the energies bounding the desired band.
Note that as expected when $E_{m,M}$ approaches 1~GeV--100~GeV the
$\sigma_K$ term vanishes, and the band-corrected flux uncertainty
reduces to that associated with the intrinsic flux uncertainty.

To obtain the associated uncertainty in the anisotropy, we make the
further assumption that the flux uncertainties among sources are
uncorrelated, and thus
\begin{equation}
\frac{\sigma_{C_P}^2}{C_P^2} 
= 
\sum_j \left( 2 w_j \F_j \right)^2 \sigma_{\F_j}^2
\bigg/
\left( \sum_j w_j \F_j^2 \right)^2\,.
\end{equation}
This ranges from 2\% at low energies to 7\% at high energies, and in
all cases, is smaller than that implied by the cosmic variance (see the
following section).

The fractional errors in $C_P$ may appear unexpectedly low given the
typical fractional flux errors of 15\%--25\% near the flux
threshold. However, the reason for this is simply that the $C_P$ is
essentially an average over $\F_j^2$.  That is, for a population of $N$
sources with identical fluxes at the threshold, this is roughly
\begin{equation}
\frac{\sigma_{C_P}^2}{C_P^2} 
\simeq
\frac{4 \F_j^2 \sigma_{\F_j}^2 N}{\F_j^4 N^2}
\simeq
\frac{4}{N} \frac{\sigma_{\F_j}^2}{\F_j^2}\,.
\end{equation}
Hence, for $N\simeq10$ sources near the threshold and
$\sigma_{\F_j}/\F_j\simeq0.25$, the 
anticipated fractional uncertainty in $C_P$ arising from the flux
uncertainty is 3\%.

\subsection{Cosmic Variance} \label{sec:CV}
Our method of summing over the unmasked 2FGL sources to obtain a lower
limit on $C_P$ (in either our conservative direct or statistical
approaches) allows a less stringent latitude cut than that applied by
the \Fermi-LAT collaboration to increase the statistics of the source
number.  Of course, this is only justified if there is no
contaminating Galactic population of point sources, which would bias
the inferred values of $C_P$.  Here, we demonstrate that this is indeed
the case: above $15^\circ$ the source populations are statistically
indistinguishable from that above $30^\circ$.  Following this, we
estimate the cosmic variance in the measured $C_P$ via the cumulative
and differential contributions to the anisotropy from various Galactic
latitude cuts.

\subsubsection{Excluding the Galactic Point-source Population}
Extending the 2FGL to lower latitudes than the $30^\circ$ cut made by
\citet{Fermi_aniso} results in a considerably larger sample size and
correspondingly lower cosmic variance.  Moreover, it provides
additional realizations with which to estimate the cosmic variance
expected for the measured anisotropy spectrum.  Here, we verify that
less stringent latitude cuts are allowed by the 2FGL (though perhaps
not by the diffuse Galactic emission) by comparing the distribution of
source properties at different latitudes via the Kolmogorov-Smirnov (K-S)
test.

Since the K-S test is one-dimensional, we explicitly compare the source
flux ($\F_{35}$) distribution and photon spectra index ($\Gamma$)
distribution separately.  A cursory comparison by eye of the joint
distribution does not reveal any noticeable correlation, suggesting
that the comparison of the projected distributions is sufficient.

The K-S test returns a probability that may be loosely interpreted as
the likelihood that two samples drawn from the same distribution
differ as much as the two being compared.  These are listed for a
variety of latitude cuts, encompassing the $30^\circ$ employed by
\citet{Fermi_aniso}, for the flux and photon spectral index
distributions in Table \ref{tab:bKS}.  (The inclusion of $b_{\rm
  min}=30^\circ$ is gratuitous and returns the expected probability
of unity.)  In these, very small values indicate differing source
populations.  Fluctuations become large at very high latitude cuts due
to the small number of remaining sources.  However, at latitudes below
$15^\circ$, the disparity can only be explained
by the presence of an additional, Galactic population.  Above
latitudes of $20^\circ$, this population is
subdominant, and above $25^\circ$ it may be
ignored altogether.  These correspond to an increase of nearly 31\%
and 14\% in the total source count, respectively.

\subsubsection{Direct Estimates of the EGRB Anisotropy for Varying
  Galactic Latitude} \label{sec:gallat} In
Figure~\ref{fig:direct_app}, we show the cumulative and
differential contribution to the EGRB anisotropy $\Delta C_P$ by
unmasked point sources for varying galactic latitude.  We perform a
Monte Carlo integration of the sky coverage after applying the
respective masks to obtain the unmasked sky fractions
$f_{\rmn{sky}}=\{0.613,0.556,0.446,0.338,0.240,0.158,0.093\}$ for
$|b|>\{5^\circ,10^\circ,20^\circ,30^\circ,40^\circ,50^\circ,60^\circ\}$
(a similar computation using the 2FGL to construct the mask yields
$f^{\rm
  2FGL}_{\rmn{sky}}=\{0.556,0.507,0.406,0.310,0.222,0.145,0.083\}$ for
the same latitude cuts). For the differential $\Delta C_P$, our
Monte Carlo integration in galactic area rings of equal area
(indicated in Figure~\ref{fig:direct_app}) yield sky fractions that
scatter around 0.09.

At low energies, there is a systematic trend of an increasing
cumulative $\Delta C_P$ for decreasing galactic latitude, which could
be due to two reasons: (1) since the different values for $\Delta C_P$
are not statistically independent (a less stringent galactic latitude
cut contains the point sources with a more conservative cut as a
subsample), this sequence of $\Delta C_P$ could be due to a regression
to a (high) mean with increasing sample size, or (2) this could sign a
population of soft-spectrum galactic point sources out to
$|b|<30^\circ$. At energies $E>10$~GeV, the differential contribution
to $C_P$ of equal-sky-area galactic latitude rings shows no systematic
trend in the (more complete) 1FHL sample and in the 2FGL sample for
$|b|>20^\circ$. The data are inconclusive whether the low-energy data
points at $|b|<30^\circ$ signal positive outliers or the hint of a
population of soft-spectrum galactic point sources. Our
K-S test suggests the first possibility and that it is
safer to use point sources with $|b|>20^\circ$ (at least at higher
energies).

\begin{figure}[ht!]
\begin{center}
\includegraphics[width=\columnwidth]{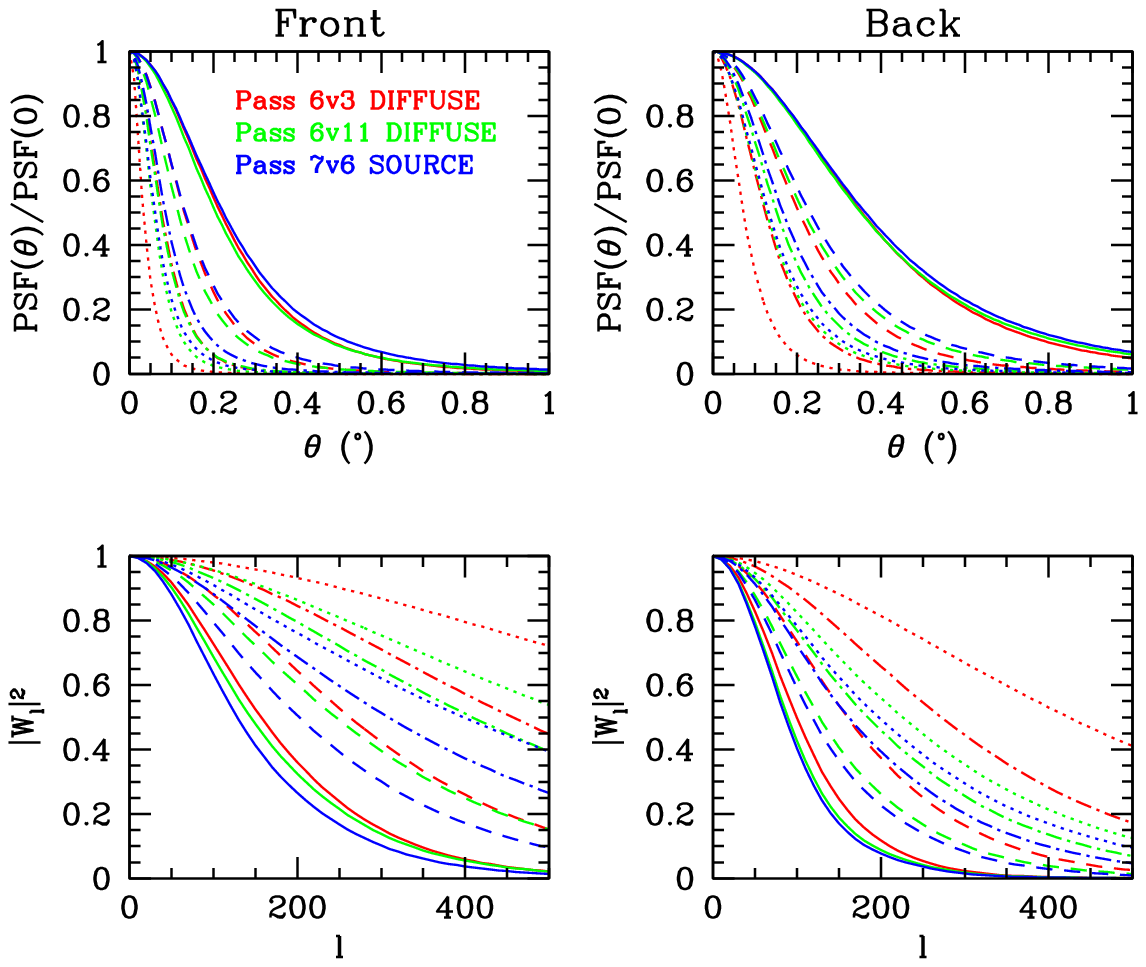}
\end{center}
\caption{Point-functions (top) for the front (left) and back
  (right) LAT detectors for the energy bands relevant for the EGRB
  anisotropy measurements for the Pass 6v3 DIFFUSE (red), Pass
    6v11 DIFFUSE (green), and Pass 7v6
  SOURCE (blue) IRFs, and their associated window functions (bottom).
  The point-spread function/window function for the 1.04--1.99~GeV,
  1.99--5.00~GeV, 5.00--10.4~GeV, and 10.4--50.0~GeV energy bands are
  shown by the solid, long-dash, short-dash, and dotted lines,
  respectively.  Varying the Pass 7v6 event class makes only a marginal
  difference.  Binning the PSF on the scales implied in
  \citet{Fermi_aniso} reduces the $|W_\ell|^2$ for the highest-energy
  bin, though it does so roughly in proportion for both
  IRFs.}\label{fig:PSF}
\end{figure}

\begin{figure}
\begin{center}
\includegraphics[width=0.75\columnwidth]{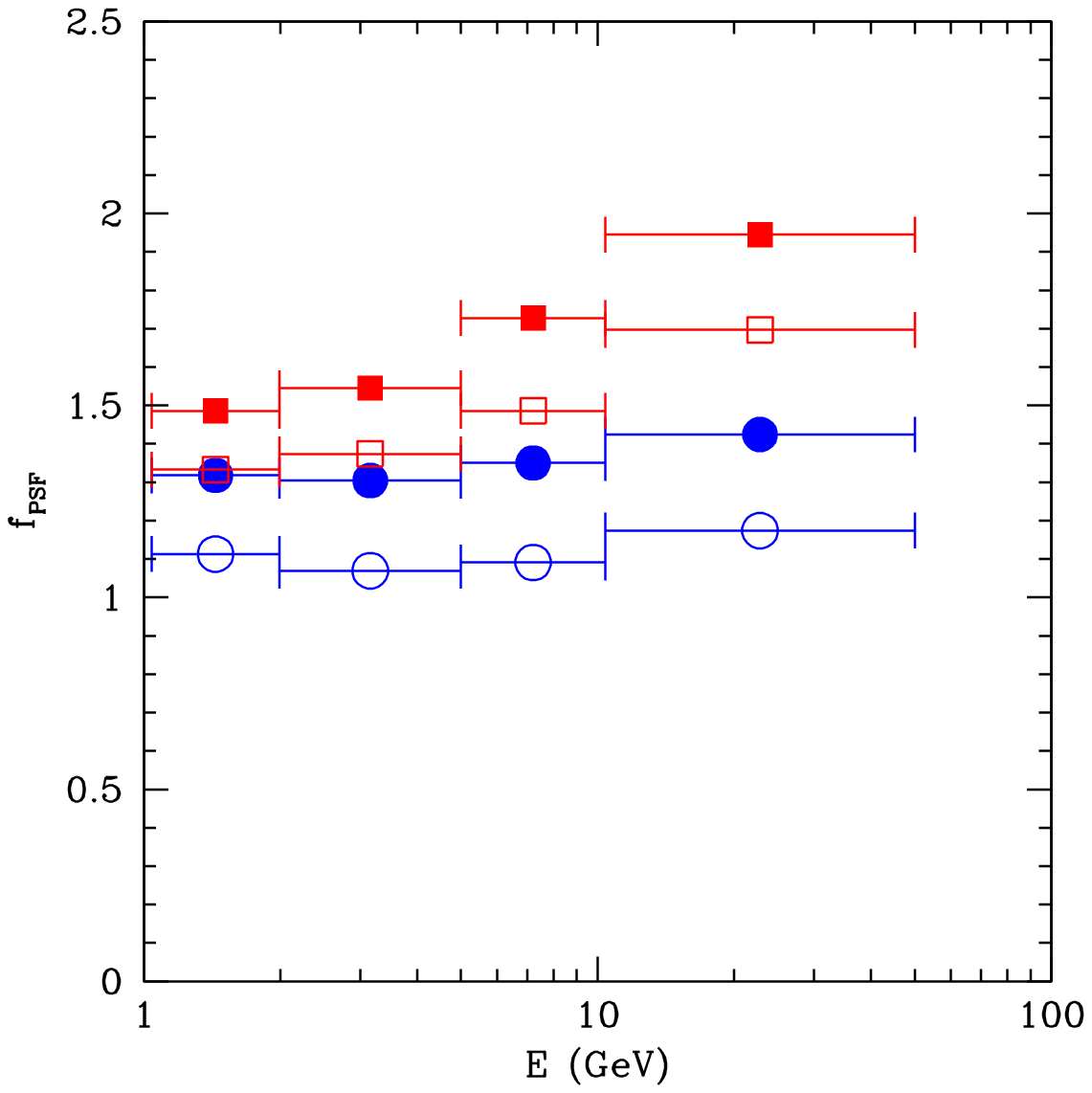}
\end{center}
\caption{Correction factor relating the EGRB anisotropy spectrum using
  the Pass 6v3 DIFFUSE and Pass 6v11 DIFFUSE IRF (open) and the
  Pass 6v3 DIFFUSE and Pass 7v6 SOURCE IRF (filled) to deconvolve
  the beam.  The correction for the front and back detector in the LAT
  are shown by the blue circles and red squares, respectively.
  Binning the PSF on the scales implied in \citet{Fermi_aniso} and
  considering alternative Pass 7v6 event classes have only a marginal
  impact.}\label{fig:fPSF}
\end{figure}

\section{Possible Sources of Reported Anisotropy Discrepancy}
Here, we collect some illustrative computations that may explain the
discrepancy between the high-energy anisotropy reported by
\citet{Fermi_aniso} and the limits implied by this analysis.  We
do not make any claims regarding completeness, restricting ourselves
to exploring plausibility.  These take two forms: modifications of the
PSF, which shift the expected measured value upward, and additional
sources of uncertainty, which lessen the statistical significance of
the discrepancy.

\subsection{\Fermi Point-spread Function} \label{sec:PSF}
We begin with a gross estimate of the impact of the recent update of
the PSF on the deconvolution procedure employed by \citet{Fermi_aniso}.
The PSF used was from the Pass 6v3 DIFFUSE IRF, and based on erroneous
Monte Carlo simulations of the detector.
However, comparisons of the apparent structure of bright, compact
Galactic and extragalactic sources conclusively demonstrated that the
Pass 6v3 PSF must be considerably broader than originally stated
\citep{Nero-Semi-Tiny-Tkac:11}.  Correspondingly, subsequent on-orbit
calibrations, first reported in Pass 6v11 IRF and used above 1~GeV in
the current IRF, Pass 7v6, exhibit much broader PSFs at high energies
for all event classes and are consistent with those found in
\citep{Nero-Semi-Tiny-Tkac:11}.  Here, we employ both the updated
Pass 7v6 and the Pass 6v11 DIFFUSE PSFs as proxies for the true PSF to
assess the potential magnitude of the correction this implies.  Note
that this is similar in principle to the comparison made in Section D of
\citet{Fermi_aniso}, though even a cursory consideration of the PSFs
shown in \citet{LAT_perf} would seem to call into question the remarkable
agreement between the Pass 6v3 and Pass 6v11 results found in
\citet{Fermi_aniso} at the highest energies.

Different IRF versions potentially place different intrinsic cuts on
individual event properties and thus may include different event
lists, complicating any direct comparison.  This is exacerbated by
changes in the stated event classes among subsequent IRF releases.
While \citet{Fermi_aniso} utilized the Pass 6v3 DIFFUSE class events,
no such event class exists in the Pass 7 IRFs.  Here, we discuss the
comparison with the Pass 7v6 SOURCE events, chosen as superficially the
most comparable based upon their description and following the
comparisons made in Section 6.6 of \citet{LAT_perf}.  However, considering
the Pass 7v6 CLEAN IRFs as a proxy instead results in only marginal
differences.

Despite the uncertainty in the event lists, it is clear that above
1~GeV, the evolution in the IRFs between Pass 6v3, Pass 6v11, and Pass 7v6 is
dominated by the differences in the PSF estimation scheme and not the
intrinsic event reconstruction \citep{LAT_perf}.  Thus, we expect that
the magnitude of the impact of the PSF on the resulting anisotropy
spectrum is well approximated by the comparison between the Pass 6v3
DIFFUSE, Pass 6v11 DIFFUSE, and Pass 7v6 SOURCE PSFs.  Differences due to the lack of the
tail term in the Pass 7 PSF parameterization will only increase the
disparity.

The PSF impacts the anisotropy measurement through the window
function, employed in the beam deconvolution and defined by
\begin{equation}
W_\ell = 2\pi \int_{-1}^1 d\mu P_\ell(\mu) {\rm PSF}(\theta)\,,
\end{equation}
where $\mu\equiv\cos\theta$ and $P_\ell$ are Legendre polynomials.  The
Pass 6v3 DIFFUSE, Pass 6v11 DIFFUSE, and Pass 7v6 SOURCE PSFs and
window functions in the energy bands of interest, assuming a spectral
index of $\Gamma=2.2$, are compared in Figure \ref{fig:PSF}.  These
may be compared with the 
window functions shown in Figure 9 of \citet{Fermi_aniso}, and
indirectly with Figure 65 of \citet{LAT_perf}.  Since the PSF
deconvolved anisotropy signal is $\propto |W_\ell|^{-2}$ \citep[see
Equation (4) of][]{Fermi_aniso}, underestimating the width of the PSF
(and thus overestimating $|W_\ell|^2$) results in 
an associated
underestimate of the anisotropy.

The magnitude of the correction to the window function depends on
the multipole at which the $C_\ell$ is determined.
Assuming that the estimate of $C_P$ arises from a least squares
fit of a constant to the observed $C_\ell$ for $150\le\ell\le500$
and that the uncertainty in the power spectrum prior to
deconvolution is dominated by photon shot noise at low $\ell$, the
relevant correction is 
\begin{multline}
f_{\rm PSF}^V 
= 
\left[\frac{\sum_\ell (2\ell+1)|W^V_\ell|^{2}}{\sum_\ell (2\ell+1)|W_\ell^V|^{4}}\right]
\bigg/
\left[\frac{\sum_\ell (2\ell+1)|W^{\rm 6v3}_\ell|^{2}}{\sum_\ell (2\ell+1)|W^{\rm 6v3}_\ell|^{4}}\right]
\,,
\end{multline}
which receives substantial contributions from large $\ell$.  This is
shown for both the back and front detectors in Figure \ref{fig:fPSF},
with a typical value at high energies ranging from 1.2 to 2.0.

In practice, the correction factor depends upon the fraction of the
anisotropy signal associated with the front and back detectors.  Given
the similarity in the effective area of the two LAT detectors,
adopting an equal anisotropy signal in each gives an effective
correction factor of roughly 1.6.  Furthermore, the anisotropy
estimates in \citet{Fermi_aniso} were obtained via the variance
weighted means of an estimate of the power spectrum separate from
front- and back-converted events, and thus, the back detector and
regions with large uncertainties are expected to contribute less to
the resulting anisotropy value.  However, at high energies, where the
discrepancies reported here are largest, the uncertainties are essentially
independent of multipole for $\ell\ge150$ and similar for the front
and back detector, reducing the impact of the variance weighting
\citep[see Section III.C of][]{Fermi_aniso}.

\subsection{$C_P$ Variance in the Few Photon Limit} \label{sec:CPvar}
The anisotropy spectrum derived in \citet{Fermi_aniso} employed
analyses derived originally for the cosmic microwave background
\citep[e.g.,][]{A55}, corresponding to the limit of small fluctuations
and large numbers of photons.  However, the statistics of power
spectrum reconstruction from a small population of dim point sources
differs significantly from that associated with many photons
continuously distributed throughout the sky. Most importantly, in the
former case, {\em source} shot noise, similar to the frequently
discussed photon shot noise, effectively limits the independence of
multipoles in the power spectrum, and thus the variance of any
reconstruction of $C_P$.  Here, we present an illustrative computation
of the variance associated with estimates of $C_P$ arising from a
population of identical point sources, computed by averaging the power
spectrum over a range in multipoles, similar in spirit to the
measurement by \citet{Fermi_aniso}.

The estimate of $C_P$ presented in \citet{Fermi_aniso} is
constructed in a number of steps, beginning with the measurement of
the $C_\ell^{\rm raw}$ from the intensity sky maps.  From these, after
correcting for the sky coverage, the photon shot noise is removed and
the PSF deconvolved, resulting in
\begin{equation}
C_\ell^{\rm signal} = \frac{C_\ell^{\rm raw}/f_{\rm sky} - C_N}{|W_\ell|^2}
\end{equation}
where $C_N$ is the average number of photons in a given sky pixel and
represents the photon shot noise term 
\citep[see Equation (4) of][]{Fermi_aniso}.
The $C_\ell^{\rm signal}$ are then averaged over 50 multipole bins,
producing an estimate of the power spectrum.  Finally, the variance
weighted mean of the power spectrum provides the desired value of
$C_P$ \citep[Equation (8) of][]{Fermi_aniso}.

Here, we will make a variety of simplifying assumptions in the interest
of clarity.\footnote{Relaxing any of these assumptions, the most
  important of which is the uniformity of the source fluxes, only
  increases the size of the corresponding source shot noise.  This is
  because doing so generally decreases the number of effective objects
  that dominate the statistics of the power spectrum reconstruction.}  
The first is that we may ignore the complications induced
by the limited sky coverage, i.e., $f_{\rm sky}=1$.  The second is
that the beam width is much smaller than the scales of interest and
thus $|W_\ell|^2\approx1$.  Together, these give 
$C_\ell^{\rm signal} = C_\ell^{\rm raw} - C_N$.  Third, we will assume
that the noise in the reconstructed power spectrum is roughly
independent of multipole ($\ell$ and $m$).  This is not strictly the
case above 5~GeV in \citet{Fermi_aniso} (see, e.g., Figure 3), for which
the variance as a function of multipole appears nearly constant
because of the suppression due to the window function.  However, the
limited range of multipoles implies that this will make at most a
factor of a few change to the variance estimates obtained here, with
the true values being somewhat larger.  With this assumption, the
photon-shot-noise variance weighted mean of the power
spectrum, the estimate of the anisotropy employed in
\citet{Fermi_aniso}, corresponds to an unweighted average over all
multipoles.  
Fourth, we will ignore any intrinsic 
source structure.  As a result, the variance weighted mean
approximately becomes an unweighted average, and the averaging within
and across bins may be done in a single step.  Therefore, the 
estimator we consider is
\begin{equation}
\begin{aligned}
C_P 
&= 
\frac{
  \sum_{\ell=\ell_\rmn{min}}^{\ell_\rmn{max}} 
  ( C_{\ell}^{\rm raw} - C_N )/\sigma_n^2
}{
  \sum_{\ell=\ell_\rmn{min}}^{\ell_\rmn{max}} 1/\sigma_n^2
}
\approx
\frac{1}{L}
\sum_{\ell,m}
C_{\ell,m}^{\rm raw} - C_N\,,
\end{aligned}
\label{eq:CPest}
\end{equation}
where we have suppressed the summation limits for compactness,
$L\equiv(\ell_\rmn{max}+1)^2-(\ell_\rmn{min}+1)^2$ is the
total number of multipoles included, $\sigma_n^2\propto (2\ell+1)^{-1}$ is
the variance due to photon shot noise alone \citep[the only source of
  noise employed by][]{Fermi_aniso}, and $\ell_\rmn{min}$ and
$\ell_\rmn{max}$ are the minimum and maximum multipoles considered.
Finally, we will assume a uniform observing 
time across the entire sky, $\Delta t$, in terms of which
\begin{equation}
C_N = \frac{n}{\Delta t^2}\,,
\end{equation}
where $n$ is the total number of events observed.

For a population of point sources, the power spectrum in the
flat-sky approximation (which is valid for $\ell_\rmn{min}\gtrsim15$
for the point-source samples considered in this work) is given by 
\begin{equation}
C_{\ell,m}^{\rm raw} = 
\frac{1}{\Delta t^2}
\sum_{a,b=1}^N M_a M_b e^{i\bmath{\ell}\cdot(\bmath{x}_a-\bmath{x}_b)}\,,
\end{equation}
where $M_{a,b}$ are Poisson deviates with mean $\bar{M}_a$ corresponding
to the expected number of photons during the entire observation for
each source, and $\bmath{x}_{a,b}$ are the positions of the sources on
the sky. The corresponding $C_P$ estimate, obtained by inserting this
into Equation (\ref{eq:CPest}), is then
\begin{equation}
C_P =\frac{1}{\Delta t^2}\left[ \frac{1}{L} \sum_{\ell,m} 
\sum_{a,b} M_a M_b e^{i\bmath{\ell}\cdot(\bmath{x}_a-\bmath{x}_b)}
-
\sum_{a} M_a
\right]
\,.
\end{equation}

The mean $C_P$ is obtained by averaging over source realizations,
i.e., both $M$ and $\bmath{x}$, which are presumed to be uncorrelated:
\begin{equation}
\begin{aligned}
\left< C_P \right>
&=
\frac{1}{\Delta t^2}
\left\{\frac{1}{L} 
\sum_{\ell,m}
\left[
\sum_{a} \left< M_a^2\right>
+
\sum_{a\ne b} \left< M_a M_b e^{i\bmath{\ell}\cdot(\bmath{x}_a-\bmath{x}_b)} \right>
\right]
-
\sum_{a} \left< M_a \right>\right\}
\\
&=
\frac{1}{\Delta t^2} \sum_a \left[ \left< M_a^2\right> - \left< M_a \right> \right]
=
\frac{1}{\Delta t^2} \sum_a \bar{M}_a^2
\end{aligned}
\end{equation}
Note this is precisely the value anticipated by Equation (\ref{eq:CP}).

The variance in this $C_P$ estimate is given by
\begin{equation}
\begin{aligned}
\sigma_{C_P}^2 
&= 
\left< C_P^2 \right> - \left< C_P \right>^2\\
&=
\frac{1}{\Delta t^4}\left\{
\frac{1}{L^2} 
\left<
\sum_{\ell,m,\ell',m'}
\sum_{a,b,c,d} M_a M_b M_c M_d 
e^{i\bmath{\ell}\cdot(\bmath{x}_a-\bmath{x}_b) + i\bmath{\ell'}\cdot(\bmath{x}_c-\bmath{x}_d)}
\right>
\right.\\
&\qquad\qquad\qquad\quad-
\frac{2}{L}
\left<
\sum_{\ell} \sum_{a,b,c} M_a M_b M_c e^{i\bmath{\ell}\cdot(\bmath{x}_a-\bmath{x}_b)}
\right>
\\
&\qquad\qquad\qquad\qquad\qquad\qquad+
\left.
\left<
\sum_{a,b} M_a M_b
\right>
-
\left(\sum_{a} \bar{M}_a^2 \right)^2
\right\}\\
&=
\frac{1}{\Delta t^4}\left\{
\frac{1}{L^2}
\left[
\sum_{(\ell,m)\ne(\ell',m')}
\left<
\sum_{a} M_a^4
+
\sum_{a\ne b} M_a^2 M_b^2
\right>
\right.\right.\\
&\qquad\qquad\qquad\qquad+\left.
\sum_{\ell,m}
\left<
\sum_a M_a^4
+
2 \sum_{a\ne b} M_a^2 M_b^2
\right>
\right]
\\
&\qquad\qquad\quad-
\frac{2}{L}
\sum_{\ell,m} 
\left<
\sum_{a} M_a^3
+
\sum_{a\ne b} M_a M_b^2 
\right>
\\
&\qquad\qquad\qquad\quad+
\left.
\left<
\sum_{a} M_a^2
+
\sum_{a\ne b} M_a M_b
\right>
-
\left(\sum_{a} \bar{M}_a^2 \right)^2
\right\}\,.\\
\end{aligned}
\end{equation}
After inserting the appropriate moments of the Poisson distribution
this simplifies to
\begin{equation}
\sigma_{C_P}^2 = \frac{1}{\Delta t^4}\left[
\sum_a 2 \bar{M}_a^2 (2\bar{M}_a+1)
+ 
\frac{2}{L} \sum_{a\ne b} \bar{M}_a (\bar{M}_a+1) \bar{M}_b (\bar{M}_b+1)
\right]\,.
\end{equation}

In addition to the extragalactic sources of primary interest here, the
background also consists of a dominant diffuse galactic component.
Above $\ell_{\rmn{min}}\approx100$, the diffuse component does not
contribute to $C_P$ significantly, though it continues to dominate
$\sigma_{C_P}^2$.  Within the context of the above formalism, we
imagine the background as due to a population of $N$ sources with
equal flux such that $\bar{M}\gtrsim1$, and $N_G\gg N$ galactic sources
with equal flux $\bar{M}_G\ll1$, in terms of which
\begin{equation}
C_P 
= 
\frac{1}{\Delta t^2} \left( N \bar{M}^2 + N_G \bar{M}_G^2 \right)
=
\frac{1}{\Delta t^2} \left( N \bar{M}^2 + \frac{n_G^2}{N_G} \right)
\end{equation}
and
\begin{equation}
\begin{aligned}
\sigma_{C_P}^2
&\approx
\frac{1}{\Delta t^4}
\left[
4 N \bar{M}^3
+
2 N_G \bar{M}_G^2
+
\frac{2}{L} \left( N^2 \bar{M}^4 + N_G^2 \bar{M}_G^2 \right)
\right]\\
&\approx
\frac{1}{\Delta t^4}
\left[
4 N \bar{M}^3
+
\frac{2 n_G^2}{N_G}
+
\frac{2}{L} \left( N^2 \bar{M}^4 + n_G^2 \right)
\right]\,,
\end{aligned}
\end{equation}
where $n_G\equiv N_G\bar{M}_G \gg n,~n_G\gg C_P\Delta t^2$, in which 
$n\equiv N\bar{M}$.  In the smooth limit for the diffuse Galactic
component $N_G\rightarrow\infty$ while $n_G$ is fixed, and thus after
some simplification, the limiting expressions are 
\begin{equation}
C_P
\approx
\frac{1}{\Delta t^2} N\bar{M}^2
\quad\text{and}\quad
\sigma_{C_P}^2
\approx
\frac{4}{n} C_P^2 + \frac{2}{L} \frac{n_G^2}{\Delta t^4}\,.
\end{equation}
The second term decreases linearly with $1/L$, i.e., as
expected for an average of independent multipoles.  However, the former
term is fixed and arises due to the correlations induced among
photons resulting from having come from a fixed number of potential
sources.  This source shot noise term, analogous to the photon shot
noise, dominates when
\begin{equation}
L\gtrsim\frac{n n_G^2}{2 C_P^2 \Delta t^4}\,.
\end{equation}

In the case of interest, roughly $N=21$ sources within the 1FHL are
capable of completely reproducing the anisotropy above 10.4~GeV, with
an average flux of 
\begin{equation}
\bar{\F} = \sqrt{\frac{C_P}{N/(4\pi f_{\rm sky})}}
\simeq
6.5\times10^{-11}~{\rm ph~\cm^{-2}~\s^{-1}}\,,
\end{equation}  
where we adopted $f_{\rm sky}=0.338$ (see Appendix \ref{sec:gallat}).
Assuming a field of view of $\Omega_{\rm fov}=2.4~{\rm sr}$, effective
area of $A_{\rm eff}=7000~\cm^{2}$, and livetime of 
$\tau_{\rm l}=56.6~{\rm Ms}$ \citep[see Section II of][]{Fermi_aniso},
this translates into roughly
\begin{equation}
\bar{M} = \bar{\F}
\frac{\Omega_{\rm fov}}{4\pi} A_{\rm eff} \tau_{\rm l}
\simeq
4.9\,,
\end{equation}
i.e., five photons per source, consistent with the reported
sensitivity in Section 2.3 of \citet{1FHL}.  Thus, the source shot 
noise effectively places a floor of $20\%$ on the fractional error in
the reconstructed value of $C_P$.  This is roughly half as large as
the reported errors above $10.4~\GeV$, suggesting that at most this
modifies the uncertainty estimates by 10\%--50\%.

\end{appendix}

\acknowledgments The authors thank Niayesh Afshordi,
Alessandro Cuoco, Jennifer Siegal-Gaskins, and the
\Fermi~collaboration for helpful discussions and Volker Springel for
careful reading of the manuscript.  A.E.B. and K.S.~receive financial
support from the Perimeter Institute for Theoretical Physics and the Natural
Sciences and Engineering Research Council of Canada through a
Discovery Grant.  Research at Perimeter Institute is supported by the
Government of Canada through Industry Canada and by the Province of
Ontario through the Ministry of Research and Innovation.
C.P.~gratefully acknowledges financial support of the Klaus Tschira
Foundation. E.P. acknowledges support by the DFG through Transregio
33. P.C. gratefully acknowledges support from the UWM Research Growth
Initiative, from \Fermi Cycle 5 through NASA grant NNX12AP24G, from
the NASA ATP program through NASA grant NNX13AH43G, and NSF grant
AST-1255469.

\bibliography{bigmh,bigmh_orig}
\bibliographystyle{apj}

\end{document}